\documentclass[sigconf]{acmart}
\usepackage[T1]{fontenc}
\def\BibTeX{{\rm B\kern-.05em{\sc i\kern-.025em b}\kern-.08emT\kern-.1667em\lower.7ex\hbox{E}\kern-.125emX}}
\usepackage{bm}
\usepackage{algorithm}
\usepackage{algpseudocode}
\usepackage{balance}

\AtBeginDocument{%
  \providecommand\BibTeX{{%
    \normalfont B\kern-0.5em{\scshape i\kern-0.25em b}\kern-0.8em\TeX}}}

\copyrightyear{2020}
\acmYear{2020}
\setcopyright{usgovmixed}
\acmConference[FPGA '20]{Proceedings of the 2020 ACM/SIGDA International Symposium on Field-Programmable Gate Arrays}{February 23--25, 2020}{Seaside, CA, USA}
\acmBooktitle{Proceedings of the 2020 ACM/SIGDA International Symposium on Field-Programmable Gate Arrays (FPGA '20), February 23--25, 2020, Seaside, CA, USA}
\acmPrice{15.00}
\acmDOI{10.1145/3373087.3375302}
\acmISBN{978-1-4503-7099-8/20/02}

\begin{document}

\fancyhead{}
\title{
Reuse Kernels or Activations? A Flexible Dataflow for Low-latency Spectral CNN Acceleration
}

\author{Yue Niu}
\affiliation{%
  \institution{University of Southern California}
  \city{Los Angeles}
  \state{CA}
  \postcode{90007}
}
\email{yueniu@usc.edu}

\author{Rajgopal Kannan}
\affiliation{%
  \institution{US Army Research Lab-West}
  \city{Los Angeles}
  \state{CA}
  \postcode{90007}
}
\email{rajgopal.kannan.civ@mail.mil}

\author{Ajitesh Srivastava}
\affiliation{
  \institution{University of Southern California}
  \city{Los Angeles}
  \state{CA}
  \postcode{90007}
}
\email{ajiteshs@usc.edu}

\author{Viktor Prasanna}
\affiliation{%
 \institution{University of Southern California}
 \city{Los Angeles}
 \state{CA}
 \postcode{90007}
}
\email{prasanna@usc.edu}

\newcommand{\cin}{c_\text{in}}
\newcommand{\cout}{c_\text{out}}
\newcommand{\hin}{h_\text{in}}
\newcommand{\win}{w_\text{in}}
\newcommand{\hout}{h_\text{out}}
\newcommand{\wout}{w_\text{out}}
\newcommand{\Xin}{\bm{X}}
\newcommand{\Yout}{\bm{Y}}
\newcommand{\Kernel}{\bm{W}}

\newcommand{\Xfreq}{\widetilde{\bm{X}}}
\newcommand{\Yfreq}{\widetilde{\bm{Y}}}
\newcommand{\Kernelfreq}{\widetilde{\bm{W}}}

\newcommand{\FFT}{\mathcal{FFT}}
\newcommand{\IFFT}{\mathcal{IFFT}}

\begin{abstract}
Spectral-domain CNNs have been shown to be more efficient than traditional spatial CNNs in terms of reducing computation complexity. However they come with a `kernel explosion' problem that, even after compression (pruning), imposes a high memory burden and off-chip bandwidth requirement for kernel access. This creates a performance gap between the potential acceleration offered by compression and actual FPGA implementation performance, especially for low-latency CNN inference. In this paper, we develop a principled approach to overcoming this performance gap and designing a low-latency, low-bandwidth, spectral sparse CNN accelerator on FPGAs. First, we analyze the bandwidth-storage tradeoff of sparse convolutional layers and locate communication bottlenecks. We then develop a dataflow for flexibly optimizing data reuse in different layers to minimize off-chip communication. Finally, we propose a novel scheduling algorithm to optimally schedule the on-chip memory access of multiple sparse kernels and minimize read conflicts. 
On a state-of-the-art FPGA platform, our design reduces data transfers by 42\% with DSP utilization  up to 90\% and achieves inference latency of 9 ms for VGG16, compared to the baseline state-of-the-art latency of 68 ms.

\end{abstract}





\maketitle

\section{Introduction}
Convolutional Neural Networks (CNNs)\cite{2012_NIPS_AlexNet,2014_arXiv_VGG,2015_CVPR_GoogLeNet,2016_CVPR_ResNet} are a popular choice for FPGA acceleration due to their widespread utility in tasks such as classification, detection\cite{2015_NIPS_FasterRNN} and segmentation\cite{2015_FCNN_CVPR}.  Accelerating CNNs on current hardware platforms brings about several major challenges: (1) model weights in the convolutional and fully-connected layers, as well as intermediate activations impose a large memory overhead; (2) convolution operations use large-scale floating-point computations, performance is thus bounded by available computing resources; (3) large data transfers between on- and off-chip memory impacts latency, throughput and total power consumption. These issues become even more critical for edge devices with only limited memory and computing capacity.


Given the large number of potentially redundant operations, compression (of the CNN model) can significantly reduce memory and computation overheads, and so is a widely accepted technique for improving efficiency. Among such methods, pruning can deliver more than $20 \times$ compression\cite{2018_ECCV_ADMMpruning}, while still maintaining high accuracy. Quantization supports traditional model operations and data access patterns, but could suffer from high accuracy drop. Another alternative is frequency domain transformation; \cite{2017_FPGA_CNNOaA,2018_FPGA_CNNCaP} convert CNNs to the spectral domain to accelerate computation without sacrificing accuracy. These come at a cost - enlarged spectral kernels and complex numbers require vastly more memory and communication bandwidth, 
causing spectral CNNs to be communication-bounded rather than computation-bounded. 
To alleviate 
these issues in spectral CNNs, \cite{2019_HiPC_SPEC2} uses an ADMM \cite{admm:boyd} based pruning method to compress spectral convolutional layers and achieve a high (uniform) compression ratio ($\sim \! 4\times$) for all kernels. 

Given these advances in compression efficiency,  the major challenge (for both spatial and spectral CNNs) now becomes addressing the {\it performance gap} - how to port these efficiency gains across to the actual hardware platform implementation. This requires careful design-space exploration and optimization. For example, while the uniform (across kernels) compression ratio of \cite{2019_HiPC_SPEC2} reduces load imbalance issues in sparse tensor operation, the problem of irregular data access still remains. This performance gap becomes especially critical for CNNs used in real-time machine learning applications on edge devices, such as face recognition\cite{2016_CMU_Openface}, autonomous driving\cite{2019_Network_Selfdriving} etc. 
In particular, computation and communication implementation overheads, unless mitigated through careful design optimizations for effectively utilizing bandwidth and minimizing latency, can severely impact performance under low-latency requirements \cite{2017_SEC_empericalLatencyStudy}


\looseness=-1 Motivated by the challenge of overcoming the performance gap between model compression and efficient low-latency hardware implementation on FPGAs,  this paper proposes a more efficient architecture for spectral convolution with sparse kernels to minimize data transfers and required bandwidth, as well as inference latency. 
We bridge this gap via a novel dataflow coupled with a (bounded, approximately-optimal) scheduling algorithm for irregular data access. We summarize our major contributions below:
\begin{itemize}\looseness=-1
    \item We provide a comprehensive complexity analysis of spectral convolutional layers, focusing on the on-chip storage versus bandwidth tradeoff. This analysis enables us to effectively identify the {\it most critical bottleneck} in each layer. 
    \item We propose a flexible dataflow for choosing the {\it optimal data reuse strategy} in different spectral convolutional layers that minimizes data transfers and off-chip communication. To support this flexibility, we also design a unified architecture that can adjust dataflow on the fly.
    \item Leveraging the fact that multiple sparse kernels access the same input, we design an {\it approximate exact-cover based scheduling algorithm} to optimally schedule on-chip memory access with minimum read conflicts. For a given a number of input image replicas, we use our schedule to  perform spectral convolution on each input tile in a minimum of clock cycles. Our scheduling algorithm assumes no pattern constraints on sparse kernels and is therefore {\it generalizable} to any set of sparse kernel inputs.
    \item We implement our design on a state-of-the-art FPGA platform. Using our flexible dataflow, data transfers are reduced by $42\%$ for VGG16. With our scheduling algorithm, DSP utilization reaches up to $90\%$ with only $10$ input replications for $64\%$ parallel kernels (on $4\times$ compressed VGG16).\looseness=-1
\end{itemize}


Compared to the baseline inference latency of 68 ms for processing all sparse spectral convolutional layers[15] (at a bandwidth of 9GB/s), our design only needs 9 ms in total with a bandwidth of 12 GB/s.  When baseline latency is scaled to match our performance, the required bandwidth becomes 58GB/s higher, far beyond single DDR capacity. Though our principal focus is on reducing latency, we are still able to achieve high throughput (112 fps).

\section{Related Work}
Model pruning\cite{2015_arXiv_DeepComp,2017_ICCV_ChannelPruning,2017_ICCV_ThiNet,2018_ECCV_ADMMpruning} aims to reduce storage for kernels, which removing redundant and insignificant connections in both convolutional and fully-connected layers. These methods either prune single connection\cite{2015_arXiv_DeepComp,2018_ECCV_ADMMpruning}, or remove a group of connections, such as the whole channel\cite{2017_ICCV_ChannelPruning}, or the whole kernel\cite{2017_ICCV_ThiNet, Lowrank_IEEE_2019}. 
Model Quantization\cite{2015_ICML_LimitedPrecision,2015_NIPS_BinaryConnect,2018_CVPR_IntegerOnly,2016_ICML_FixpointQuant} compresses models by encoding float-point weights or activations into short-bit representations. Quantization can reduce computation complexity by converting float-point operation into integer or bit operation, while still holding regular dataflow that is friendly to hardware implementation.

Besides model compression, hardware implementation also assumes a key role in accelerating CNNs' deployment. Current hardware platforms either fail to efficiently parallelize large-scale convolution/matrix operations\cite{2016_MICRO_Knights}, or have a fixed memory and computing hierarchy which cannot fit all CNN models\cite{2018_MICRO_Volta}. 
Facing these limitations in current hardware platforms, domain-specific architecture emerges as an effective solution for accelerating CNN inference. FPGA based architectures\cite{2016_FPGA_SVD,2018_TCAD_Caffine, Lowrank_GlobalSIP_2017,2017_FPGA_CNNOaA,2018_FPGA_CNNCaP} achieve very high throughput or low latency by tailoring its configurable resources (DSPs/BRAMs/LUTs) to certain model. 
In this way, each model can be quickly deployed into FPGA platforms. Along with design space exploration, performance and resources can be well-balanced. 
Among these works, \cite{2016_FPGA_SVD} adopts kernel-shaped PE array which processes all operations in one kernel simultaneously. High throughput on targeting model (VGG16) can be obtained, but at the cost of less configurability for different model with different kernel sizes. 
\cite{2018_TCAD_Caffine} views convolutional and fully-connected layers as uniformed matrix-multiplication process. In this way, a unified micro-architecture is proposed, in which dataflow for convolutional and fully-connected layers can be separately optimized to maximize FPGA computing and bandwidth utilization. 
Being aware of a lot of data reuses in convolutional layers, \cite{2017_DAC_CNNSystolic} proposes a systolic-array architecture which fully exploit data reuses by enabling communication among multiple processing elements (PEs). 

Different from previous works, \cite{2017_FPGA_CNNOaA} \cite{2018_FPGA_CNNCaP} convert spatial convolution into spectral domain. 
By first tiling input images, then doing Fast Fourier Transform (FFT) on each input tile, they convert computational-intensive convolution operations into light-weighted Hadamard operations in spectral domain.
The advantage is two fold. First, data flow in convolutional layers is being simplified with Hadamard product. Then, computation complexity is reduced without sacrificing accuracy. 
For example, for VGG16, computation complexity can be reduced by $3\times$ under input tile size of $8\times 8$. However, these two methods come with the serious issue of kernel memory explosion in spectral domain. 
For VGG16, by converting $3\times 3$ spatial kernel to $8\times 8$ complex-number kernel in spectral domain, storage increases by almost $15\times$. 
This issue become even more urgent if we use larger FFT size, which cause convolutional layer to be memory-bounded instead of computation-bounded. As a result, reduction in computation complexity cannot be fully exploited in hardware platforms due to high memory and communication overhead.


\section{Background}
\textbf{Spatial CNNs}, as we call, are convolutional neural networks that consist stacked conventional convolutional layers, followed by activation functions (like ReLU), then usually end with fully-connected layers. 
Convolutional layers are used to abstract features from inputs and combine them to further higher-level representations\cite{2014_ECCV_UnderstandCNN}. 
Given input activations $\Xin\in\mathbb{R}^{b\times\cin\times\hin\times\win}$, and kernels $\Kernel\in\mathbb{R}^{\cout\times\cin\times k\times k}$, outputs $\Yout\in\mathbb{R}^{b\times\cout\times\hout\times\wout}$ can be calculated as:
\begin{equation}\label{eq: SpatialConv}
    \Yout = \Big \{\Yout_{i,n} \Big \}=\Big \{\sum_{m=1}^{\cin} \Xin_{i,m} * \Kernel_{n,m} \mid 1\leq i\leq b, 1\leq n \leq \cout \Big \},
\end{equation}
where $b$ denotes the batch size, $\cin$ and $\cout$ denotes \#channels for input and output activations. $\hin$,$\win$,$\hout$ and $\wout$ indicate spatial dimension of input and output activations. ``$*$'' is the convolution operator\footnote{Adding bias is omitted in our analysis.}.

Fully-connected layers only consist normal matrix-vector operations with flattened input batch $\Xin\in\mathbb{R}^{b\times M}$, which compute output activations $\Yout\in\mathbb{R}^{b\times N}$. Given kernel $\Kernel\in\mathbb{R}^{M\times N}$, $\Yout$ is obtained as show in Eq~(\ref{eq: SpatialFC}).
\begin{equation}\label{eq: SpatialFC}
    \Yout = \Xin \times \Kernel,
\end{equation}
where $M$ and $N$ denotes the length of input and output activations.

Compared with fully-connected layers, convolutional layers account for most computations. For example, in VGG16, $99\%$ computations fall into convolutional layers. On the other side, convolutional layers only have $10\%$ of total parameters, which brings spatial convolutional layers to be bounded by available computing resources.

\textbf{Spectral CNNs} are the models that convert convolution in spatial domain into Hadamard product in spectral domain (frequency domain) with Fast Fourier Transform (FFT)\cite{2017_FPGA_CNNOaA}. 
To reduce complexity of FFT and simplify hardware implementation, it first breaks up inputs $\Xin$ into small tiles $\Xin^{t,s}\in\mathbb{R}^{b\times \cin \times \hin^{'} \times\win^{'}}$, where $t,s$ denote the position in the whole image for each single tile, $\hin{'},\win{'}$ is tile size. Usually $\hin{'}$ equals $\win{'}$.
Then given window size $K=\hin{'}+k-1$, it applies FFT on these small tiles, getting the spectral-domain counterpart $\Xfreq^{t,s}\in\mathbb{R}^{b\times\cin \times K \times K}$. 
After applying spectral kernels $\Kernelfreq\in\mathbb{R}^{\cout\times\cin\times K \times K}$ on each tile, and accumulating along the dimension of input channels, we get spectral-domain output tiles $\Yfreq^{t,s}\in\mathbb{R}^{b\times\cout\times K \times K}$, as shown in Eq~(\ref{eq: SpectralConv}).
\begin{equation}\label{eq: SpectralConv}
    \Yfreq^{t,s} = \Big\{\Yfreq^{t,s}_{i,n} \Big \} = \Big \{\sum_{m=1}^{\cin} \Xfreq^{t,s}_{i,m} \circ \Kernelfreq_{n,m} \mid 1\leq i\leq b, 1\leq n \leq \cout \Big \},
\end{equation}
where ``$\circ$'' denotes Hadamard product.

Finally, we use Invert Fast Fourier Transform (IFFT) to convert these tiles back to spatial domain, getting $\Yout^{t,s}\in\mathbb{R}^{b\times\cout\times K \times K}$. 
To get spatial-domain activations for further non-linear operation (like ReLU, Pooling), we need to concatenate these tiles.
Due to the sliding window operation of convolution, these output tile $\Yout^{t,s}$ have some overlaps on the boundary. Hence, an operation called Overlap-and-Add (OaA) operation is applied to add and merge the overlap regions\cite{2017_FPGA_CNNOaA}.

\begin{equation}\label{eq: IFFT}
    \Yout = \bigcup_{t,s} \Yout^{t,s} = \bigcup_{t,s} \IFFT(\Yfreq^{t,s}),
\end{equation}
where ``$\bigcup$'' denotes operations that add and merge output tiles in the same image.

Converting CNNs into spectral domain reduces computation complexity in most convolutional layers. 
For VGG16, with FFT size $K=8$, computation complexity can be reduce by $2\times$ without any accuracy degradation\cite{2017_FPGA_CNNOaA}. Furthermore, Hadamard operation simplifies the dataflow in which sliding window on each input tiles can be avoided.

\textbf{FPGA Acceleration} for CNNs takes advantage of FPGAs to parallelize computations in CNNs and customize corresponding dataflow. 
Due to intrinsic parallel architecture in FPGA, convolution and matrix operations can be efficiently mapped into DSP arrays on FPGAs. 
The main focus on FPGA acceleration is to customize memory hierarchy so that data movement can be minimized between on- and off-chip memory and maximized in local memory buffer. 
Doing this, PEs can be guaranteed to reach peak performance during computing. 
Unlike ASICs, required on-chip memory in FPGA is mapped into the array of Block RAMs (BRAMs), in which each BRAM can store 36Kb data\footnote{36Kb BRAM is the basic unit in Xilinx devices, for Intel FPGA, it will be 20Kb}. 
The logic of controlling dataflow and computing is mapped into Look-Up Table (LUT) arrays in FPGA. Each LUT can be configured as various gate function. 
Hence, given available resources above ($N_{DSP}$ DSPs, $N_{BRAM}$ BRAMs, $N_{LUT}$ LUTs, as well as off-chip bandwidth $BW_{sys}$), FPGA designs are aimed to optimize certain objective function $T$ under these resource constraints
\begin{equation}\label{eq: FPGADesign}
\begin{aligned}
    & \text{opt} & T & \\
    & \text{subject to} & n_{DSP} \leq N_{DSP} & \\
    & & n_{BRAM} \leq N_{BRAM} & \\
    & & n_{LUT} \leq N_{LUT} & \\
    & & bw \leq BW_{sys}
\end{aligned}
\end{equation}
where $n_{DSP},n_{BRAM}, n_{LUT}$ and $bw$ are the required resources in the design. The objective function $T$ can be to maximize throughput, minimizing latency, or minimizing power consumption, etc.

Nowadays, since CNN compression techniques (pruning and quantization) are playing a key role in edge devices, FPGAs become even more important due to its powerful configurability. 
For example, pruned models require dedicated dataflow to address irregular data access and load imbalance, which makes the FPGA be a perfect platform to design specialized memory hierarchy to fully exercise potentials in compressed CNNs. 

\section{Complexity Analysis}\label{sec::analysis}
Given enlarged spectral kernels, even in compressed spectral CNNs, memory and communication have totally different patterns compared with spatial CNNs. 
In edge devices, due to limited on-chip memory and bandwidth, also low-latency requirement, trade-off between buffering data and off-chip communication becomes nontrivial. 
Hence conducting a complexity analysis can help locate the primary bottleneck and explore new trade-offs in each spectral convolutional layer.

The complexity analysis is based on a general architecture model, in which data (inputs, outputs, kernels) are originally stored in off-chip memory, then streamed into on-chip memory. 
For each spectral convolutional layer, computing engine first fetches tiles of inputs and kernels, then computes Hadamard product and accumulation, finally writes output tiles to off-chip memory, as shown in Fig.~\ref{fig::arch_overview}. 
In the analysis, we assume each $K\times K$ spectral kernel is compressed by $\alpha$, in which only $\frac{K^2}{\alpha}$ values are non-zeros. 
In this context, we analyse the complexity from two aspects: On-chip storage, communication bandwidth. 
\subsection{On-chip storage}\label{sec::analysis::storage}
In most cases, on-chip memory resources are not capable of keeping all kernels/activations. 
The common technique is to tile kernels/activations, calculating partial outputs each time. In next round, we keep some kernels/activations on chip to be reused and bring some new data. 
Hence, different data reuse approaches require different number of BRAMs. In spectral convolutional layers, we can choose to reuse input tiles/kernels/output partial sums. 

To reuse input tiles, we fix input activations on chip until they serve all kernels, while kernels are grouped to be fed into on-chip buffers. 
In doing this, since kernel buffers are overwritten by another group of kernels, we have to re-load them when starting operation on new input tiles. 
On the contrary, reusing kernels leads to an opposite dataflow in which parts of kernels are kept on chip, streaming input tiles instead. 
Reusing output partial sums is to keep partial sums on chip until all input channels of current tile are finished, which avoids writing and re-loading partial sums back and forth.

To give a general analysis, suppose we parallelize $N^{'}$ kernels, $M^{'}$ input channels, and $P^{'}$ input tiles, given the fact that each BRAM can only support one concurrent access, each parallel line (kernel, input channels, input tiles) needs at least one BRAM to support parallel accesses. 
Besides, we also need to write partial sums to output buffer in parallel. 
In addition, if kernels are sparse, we need some replicas ($r$) for each input tile to support accesses to the same input tile from multiple kernels (See Sec.~\ref{sec::arch::scheduling}).

In the case of reusing kernels and output partial sums (\textbf{Flow \#1}), we keep kernels and partial sums on chip, streaming all input tiles, the required BRAMs will be at least
\begin{equation}\label{eq: reqBRAMs1}
    n_{BRAM} = \underbrace{rM^{'}P^{'}}_{\text{Input tiles}} + \underbrace{M^{'}N^{'}}_{\text{Kernels}} + \underbrace{N^{'}P^{'}\left \lceil \frac{\hin\win\cdot K^2}{P^{'}\hin{'}\win{'}\cdot 1024} \right \rceil}_{\text{partial sums}},
\end{equation}
where ``$1024$'' indicates memory depth for single BRAM, and ``$K^2$'' is output tile size.

On the other side, if we choose to reuse input tiles and output partial sums, streaming kernels (\textbf{Flow \#2}), the required BRAMs will be at least
\begin{equation}\label{eq: reqBRAMs2}
    n_{BRAM} = \underbrace{rM^{'}P^{'}}_{\text{Input tiles}} + \underbrace{M^{'}N^{'}}_{\text{Kernels}} + \underbrace{M^{'}P^{'}\left \lceil \frac{N\cdot K^2}{N^{'}\cdot 1024} \right \rceil}_{\text{partial sums}},
\end{equation}

If we reuse input tiles and kernels, streaming partial sums (\textbf{Flow \#3}), $n_{BRAM}$ can be obtained as follows:
\begin{equation}\label{eq: reqBRAMs3}
\begin{aligned}
    n_{BRAM} = \min & \left \{  \right. \underbrace{rM^{'}P^{'}\left \lceil \frac{\hin\win\cdot K^2}{P^{'}\hin{'}\win{'}\cdot 1024} \right \rceil}_{\text{Input tiles}} + \underbrace{M^{'}N^{'}}_{\text{Kernels}} + \underbrace{M^{'}P^{'}}_{\text{partial sums}}, \\
    & \underbrace{rM^{'}P^{'}}_{\text{Input tiles}} + \underbrace{M^{'}N^{'}\left \lceil \frac{\frac{1}{\alpha}\cdot N\cdot K^2}{N^{'}\cdot 1024} \right \rceil}_{\text{Kernels}} + \underbrace{M^{'}P^{'}}_{\text{partial sums}} \left.  \right \},
\end{aligned}
\end{equation}
where $\alpha$ is the compression ratio. 

For some layers, $n_{BRAM}$ can be very large in certain flow, even beyond the system capacity, which are detailed in Sec \ref{sec::arch::flexible_flow}. Therefore, through this analysis, we can accurately describe if certain flow is bounded by BRAMs. On the other hand, we can also find out which flow needs minimum number of BRAMs.

\subsection{Communication bandwidth}\label{sec::analysis::bandwidth}
Different dataflows bring different communication overhead between on- and off-chip memory. 
Suppose for certain spectral convolutional layer $i$, the latency is set to be $\tau_i$ (s),the lower bound of the required bandwidth is: $bw = \frac{\#\text{Data transfers}}{\tau_i}$. 
Off-chip communication consists of reading inputs and kernels, writing outputs, writing and re-loading partial sums. 
Given three dataflow above, we formulate the required bandwidth as follows:

In \textbf{Flow \#1}, input buffer is overwritten by new input tiles after each round. If we start computation with new group of kernels, we have to \textit{re-load} corresponding input tiles. 
Given latency $\tau_i$ in layer $i$, the required bandwidth is:
\begin{equation}\label{eq: reqBW1}
    bw = \underbrace{\frac{M\hin\win\cdot\frac{N}{N^{'}}}{\tau_i}}_{\text{Inputs}} + \underbrace{\frac{\frac{1}{\alpha} \cdot NMK^2}{\tau_i}}_{\text{Kernels}} + \underbrace{\frac{N\hout\wout}{\tau_i}}_{\text{Outputs}},
\end{equation}
where ``$\frac{N}{N^{'}}$'' means the number of times for re-loading input tiles.

In \textbf{Flow \#2}, kernel buffer is overwritten instead. Compared with \textbf{Flow \#1}, the main communication rests on re-loading kernels.

\begin{equation}\label{eq: reqBW2}
    bw = \underbrace{\frac{M\hin\win}{\tau_i}}_{\text{Inputs}} + \underbrace{\frac{\frac{1}{\alpha}\cdot NMK^2\cdot \frac{\hin\win}{P^{'}\hin{'}\win{'}}}{\tau_i}}_{\text{Kernels}} + \underbrace{\frac{N\hout\wout}{\tau_i}}_{\text{Outputs}},
\end{equation}
where ``$\frac{\hin\win}{P^{'}\hin{'}\win{'}}$'' means the number of times for re-loading kernels.

Different with \textbf{Flow \#1} and \textbf{Flow \#2}, \textbf{Flow \#3} needs to write partial sums to off-chip memory and re-load them if needed, which stands out as the main communication overhead.
\begin{equation}\label{eq: reqBW3}
    bw = \underbrace{\frac{M\hin\win}{\tau_i}}_{\text{Inputs}} + \underbrace{\frac{\frac{1}{\alpha}\cdot NMK^2}{\tau_i}}_{\text{Kernels}} + \underbrace{\frac{N\hout\wout\cdot2\cdot\frac{M}{M^{'}}}{\tau_i}}_{\text{Outputs}},
\end{equation}
where ``$2\cdot\frac{M}{M^{'}}$'' means the number of times for writing/reading partial sums.

Combining these two analyses, each layer's overhead of both on-chip storage and communication for each dataflow is brought to the surface. 
Knowing the overhead, we can find an optimal dataflow, even a hybrid one which combines different flow strategies, to minimize hardware overhead, as detailed in Sec.~\ref{sec::arch::flexible_flow}.

\section{Architecture Design}\label{sec::arch}
\subsection{Overview}
\begin{figure*}[!htb]
    \centering
    \includegraphics[width=.8\textwidth]{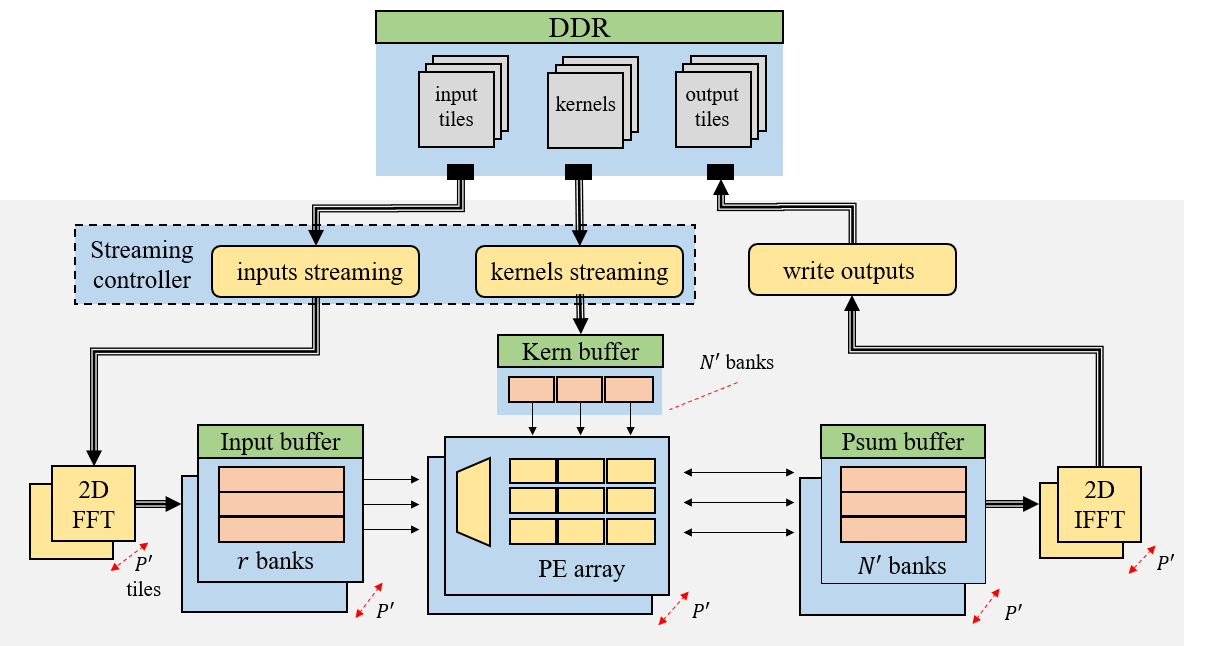}
    \caption{Overview of spectral CNN Arch.}
    \vspace{-.4cm}
    \label{fig::arch_overview}
\end{figure*}

According to previous analysis, each spectral convolutional layer might prefer certain dataflow to make an optimal trade-off between memory and communication overhead. 
To support this flexibility, we design a unified architecture that can adjust dataflow on the fly, as shown in Fig.~\ref{fig::arch_overview}. 
Streaming controller decides what data (inputs, kernels, partial sums) should be reused or streamed based on each layer's configurations (Sec.~\ref{sec::arch::flexible_flow}). 
Also, irregular data access can be well-solved by joint optimization of scheduling algorithm and supportive hardware implementation (Sec.~\ref{sec::arch::scheduling}).

The whole process is: all inputs, kernels, and outputs are originally stored in DDR memory; 
once FPGA kernel starts, input tiles and kernels are fed into on-chip buffer before enabling process elements (PEs); 
then, these tiles are converted into spectral domain by 2D FFT (kernels are already in spectral representation); 
then we start all parallel PEs, calculating Hadamard product and accumulating partial sums along input channels; 
after all input channels of current tiles are done, output tiles are converted back to spatial domain using 2D IFFT before writing into DDR memory. 
To reduce latency, we choose to process multiple input tiles and kernels simultaneously, and process input channels in a serial manner ($M^{'}=1$) so that write conflicts can be avoided. 
In each round, one input tile can be reused by all $N^{'}$ parallel kernels, while each kernel can be reused by $P^{'}$ tiles.

\subsection{Flexible dataflow}\label{sec::arch::flexible_flow}
\begin{figure*}[!htb]
    \centering
    \includegraphics[width=0.8\textwidth]{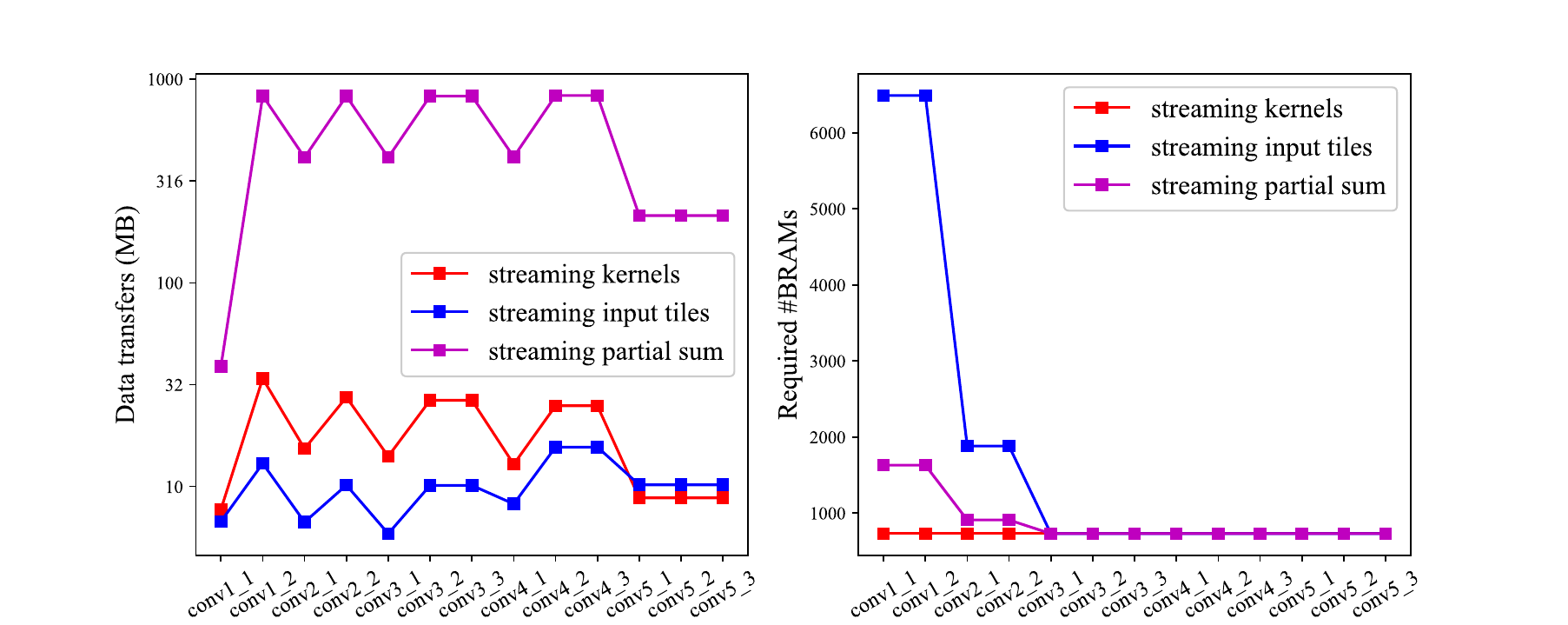}
    \caption{Complexity in VGG16(Left: Data transfers; Right: Required BRAMs)}
    \vspace{-.4cm}
    \label{fig::complexity_VGG16}
\end{figure*}

Keeping a fixed dataflow across all layers cannot guarantee data transfers are minimized. 
Based on the analysis in Sec.~\ref{sec::analysis}, the required on-chip memory and communication bandwidth differs among dataflows. For example, Fig.~\ref{fig::complexity_VGG16} shows required on-chip memory and off-chip bandwidth in three dataflows for compressed VGG16 ($\alpha=4$): \textbf{Flow \#1} (streaming input tiles), \textbf{Flow \#2} (streaming kernels), and \textbf{Flow \#3} (streaming partial sums). 

It is easy to see that streaming input tiles (reuse kernels and partial sums) though has fewer data transfers, it requires huge number of BRAMs. 
On the other side, streaming kernels needs fewer BRAMs but poses higher communication overhead. 
Another interesting thing comes with streaming partial sums which brings no advantages at all due to its huge amount of data transfers. 
Similar observations also hold in other compressed networks. 
Inspired by this, we propose an optimization algorithm to flexibly decide \textit{\#kernels should be processed before flushing corresponding input tiles, and \#input tiles before flushing corresponding kernels}. We call these two parameters \textbf{streaming parameters}: $Ns$ (kernels) and $Ps$ (input tiles). Then the required BRAMs is at least:
\begin{equation}\label{eq::reqBRAMsflexbile}
    n_{BRAM} = \underbrace{rP^{'}}_{\text{Input tiles}} + \underbrace{N^{'}\left \lceil \frac{\frac{1}{\alpha}\cdot N_sK^2}{N^{'}\cdot 1024} \right \rceil}_{\text{Kernels}} + \underbrace{N^{'}P^{'}\left \lceil \frac{N_sP_sK^2}{N^{'}P^{'}\cdot1024} \right \rceil}_{\text{partial sums}}.
\end{equation}
The required bandwidth is at least:
\begin{equation}\label{eq::reqBWflexible}
    bw = \underbrace{\frac{M\hin\win\cdot\frac{N}{N_{s}}}{\tau_i}}_{\text{Inputs}} + \underbrace{\frac{\frac{1}{\alpha}NMK^2 \cdot \frac{\hin\win}{P_s\hin^{'}\win^{'}}}{\tau_i}}_{\text{Kernels}} + \underbrace{\frac{N\hout\wout}{\tau_i}}_{\text{Outputs}}.
\end{equation}
\begin{algorithm}
\caption{Dataflow optimization algorithm}
\label{algo::flowOpt}
\begin{algorithmic}[1]
\renewcommand{\algorithmicrequire}{\textbf{Input:}}
\renewcommand{\algorithmicensure}{\textbf{Output:}}
\Require Compressed CNN model with compress ratio $\alpha$.
\Ensure \#parallel input tiles $P^{'}$; \#parallel kernels $N^{'}$; streaming parameters $Ps$, $Ns$;
\For{($P^{'}$,$N^{'}$) \textbf{in} \text{possible arch parameters}}
\For{$lyr$ \textbf{in} \text{all conv layers}}
    \State Get layer parameter ($M$, $N$, $P$,  $\hin$,$\win$, $K$, $\hin^{'}$,$\win^{'}$)
    \For{($Ps$, $Ns$) \textbf{in} \text{all possible streaming parameters}}
        \State $n_{BRAM1} \gets \text{\#BRAMs needed for } \textbf{Flow \#1}$
        \State $n_{BRAM2} \gets \text{\#BRAMs needed for } \textbf{Flow \#2}$
        \State $n_{BRAM3} \gets \text{\#BRAMs needed for } \textbf{Flow \#3}$
        \State \text{Choose the flow with minimum \#BRAMs} 
        \State $n_{BRAM}\gets \min(n_{BRAM1}, n_{BRAM2}, n_{BRAM3})$
        \If{$n_{BRAM} < N_{BRAM}$}
        \State $bw \gets \text{Calculate bandwidth (Eq.~\ref{eq::reqBWflexible})}$
        \If{$bw < bw_{\text{min}}$}{\color{blue}\Comment{Current streaming}}
            \State $bw_{\text{min}} \gets bw$
            \State Update steaming parameter ($Ps$, $Ns$)
        \EndIf
        \EndIf
    \EndFor
\State $bw_{\text{max}} \gets \text{Register max bandwidth in all layers}$
\EndFor
\If{$bw_{\text{max}} < bw_{\text{arch}}$}{\color{blue}\Comment{Current arch}}
\State $bw_{\text{arch}} \gets bw_\text{max}$
\State Update arch parameter ($P^{'}, N^{'}$)
\EndIf
\EndFor
\end{algorithmic} 
\end{algorithm}
\vspace{-.2cm}

Considering different \textbf{architecture parameters} ($P^{'}$, $N^{'}$) also lead to different performance, to get the optimal trade-off, we design a heuristic optimization method to find optimal settings for both architecture and streaming parameters, as shown in Alg.~\ref{algo::flowOpt}.

Given a compressed model, Alg.~\ref{algo::flowOpt} does a heuristic search in architecture parameter space, as well as in streaming parameter space. 
In each layer, we fix architecture parameters ($P^{'}$, $N^{'}$), trying all possible streaming parameters ($Ps$, $Ns$). 
In each streaming setting, we choose the one with minimum required BRAMs, then calculate its bandwidth $bw$. 
We update the optimal streaming parameter if the resulting bandwidth $bw$ is less than previous minimum $bw_{\text{min}}$. 
After iterating all layers, we register the maximum bandwidth $bw_{\text{max}}$ as bandwidth requirement in current architecture setting. 
Finally, we choose the pair ($P^{'}$, $N^{'}$) and ($Ps$, $Ns$) with minimum $bw_{\text{max}}$ given a network model.

\begin{figure}[!htb]
    \centering
    \includegraphics[width=0.4\textwidth]{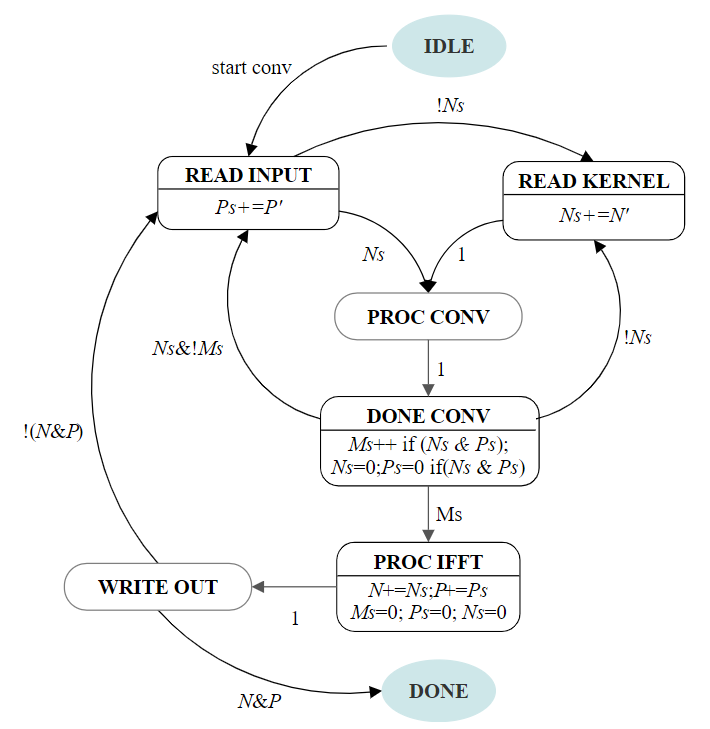}
    \caption{Streaming controller}
    \vspace{-.4cm}
    \label{fig::streamingcontrol}
\end{figure}

Streaming controller in Fig.~\ref{fig::arch_overview} is designed to adjust dataflow in different layers. 
Streaming options are managed by a internal state machine, as shown in Fig.~\ref{fig::streamingcontrol}. 
Each time when spectral convolution is done (\textbf{DONE CONV}), the state machine first checks if the number of processed kernel reaches $N_s$. 
If not ($!N_s$), it continue reading kernels (\textbf{READ KERNEL}). Otherwise, when all input channels of current kernels haven't been done ($!M_s$), two cases can arise: first, $P_s$ input image tiles have been finished, for which we need to flush all kernels and input tiles, loading input tiles and kernels from new input channel; 
second, we are still on the way of processing $P_s$ input tiles in current input channel, for which we need to load new input tiles, while kernels are already loaded. 
Another situation is that $M_s$ input channels for current $N_s$ kernels and $P_s$ tiles are done, it then starts IFFT operation (\textbf{RPOC IFFT}) to convert output image tiles back into spatial domain, then write to off-chip memory (\textbf{WRITE OUT}). 
At this time, we check if all kernels and input tiles are done, if not ($!(N \& P)$), we go back to read input tiles and kernels, otherwise, we complete current spectral convolutional layer (\textbf{DONE}).

\subsection{Memory access scheduling}\label{sec::arch::scheduling}
\begin{figure}[!htb]
    \centering
    \includegraphics[width=0.4\textwidth]{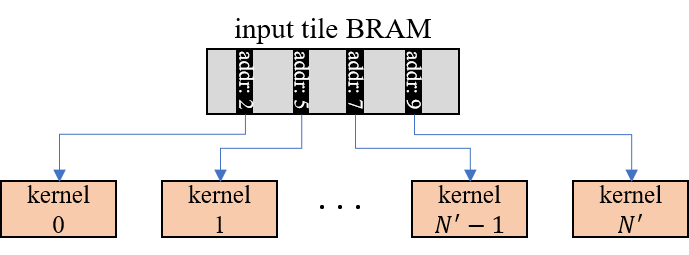}
    \caption{Sparse kernel access}
    \vspace{-.4cm}
    \label{fig::kernel_access}
\end{figure}

Memory access scheduling comes with the fact that parallel sparse kernels assume different access patterns to input tiles. As shown in Fig.~\ref{fig::kernel_access} which is a typical case of memory access in our design, in which multiple kernels are processed in parallel, while weight values in each kernel are streamed in serial. 
In each cycle, each PE sends a read address to the input BRAM. After getting the input, PE multiplies it with a corresponding kernel value, then accumulate with the one in partial sums buffer (Fig.~\ref{fig::arch_overview}). Given $N^{'}$ parallel kernels, at worst, one input BRAM needs to provide $N^{'}$ different values in a single clock cycle, which violates BRAM's principle of single access.

To prevent multiple kernels from starving for no available inputs, we use $r$ replicas to increase throughput of input BRAMs. 
And noting that some kernels might access the same location, which can be served by one BRAM, the number of replicas can be smaller than $N^{'}$. 
Given $N^{'}$ sparse kernels and $r$ replicas for each input tile, in order to minimize clock cycles of Hadamard product on current $N^{'}$ kernels and $P^{'}$ tiles, we design a novel scheduling algorithm to optimally group read requests.

Our scheduling arises from the notion that the order of processing each value in a kernel does not matter. 
As long as the corresponding indices come along with the value to specify where to write corresponding result into, we can rearrange values in current kernel group in any order. 
The scheduling algorithm is aimed to \textit{group access addresses so that the number of distinct addresses in each read cycle is less than $r$, while the number of cycles of finishing current group of sparse kernels reach minimum}.
Given $N^{'}$ sparse kernels, in which each kernel is represented by a format of $\left \{ (val, index) \mid 0\leq index \leq K^2 \right \}$. 
The number of indices in each kernel exactly equals to $\frac{K^2}{\alpha}$. 
Therefore, the scheduling can be obtained by solving the following problem: given a matrix $M$ of size $N' \times K^2/\alpha$ where each row stores indices of the same kernel, rearrange values in each row to minimize $\sum_{i=1}^{K^2/\alpha} \lceil ID_i/r \rceil$, where $ID_i$ is the number of distinct indices in the $i^{th}$ column. 
we refrain from rearranging within the columns as it may result in memory conflicts in the output. 

\begin{figure}[!htb]
    \centering
    \includegraphics[width=0.3\textwidth]{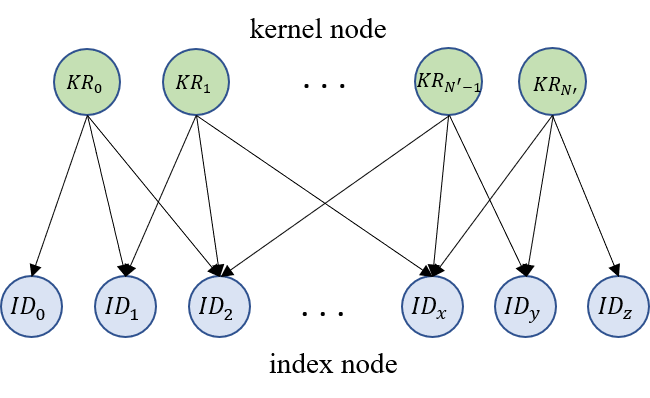}
    \caption{Bipartite presentation of sparse kernels}
    \vspace{-.4cm}
    \label{fig::kernelbipartite}
\end{figure}

We first transform the representation into a bipartite graph, as shown in Fig.~{\ref{fig::kernelbipartite}}, in which each connection between kernel node and index node means the kernel has non-zero value in corresponding index $ID_x$. 

Based on this bipartite graph, we can construct a unit scheduling set $s_i = \left \{ (KR_{x_i}, ID_{x_i}),...,(KR_{y_i}, ID_{y_i})\right \}$ that meets following constraints: 
(\textbf{C1}) elements in each set have distinct kernel nodes $KR_x$, which reflects the fact that only one value in each kernel can be processed in each cycle; 
(\textbf{C2}) the number of different indices $ID_x$ cannot exceed $r$, which enables $r$ replicas to feed all selected kernels. 
Each set $s_i$ represents one-cycle read access of $N^{'}$ parallel PEs. Since kernels are shared among $P^{'}$ input tiles, $s_i$ can be broadcast to all $P^{'}$ input tiles.
Ideally, we can find all welcome sets $S=\left \{s_1, s_2, ..., s_i, ...\right \}$. 
Then, to get an optimal scheduling which has minimum clock cycles, we need to find the set collection $S^{*} \subset S$ which has the minimum number of sets. 
This is exactly an \textbf{exact cover} problem. 

Given huge number of possible set $s_i$ and exact-cover problem is NP-complete, we choose to use a greedy algorithm to efficiently approximate the optimal solution, as shown in Alg.~\ref{algo::sheduling}. 
In each step, we find a set $s_i$ that greedily leads to optimal solution. 
Two cases come up during searching: First, if there are a set collection $S^{'}$ covering all kernel nodes, then we need to decide which sets to be chosen so that it causes minimal adverse effects on following optimization. 
We choose to leave as many high-degree index nodes as possible for future search since it's easier to find a set that can cover most kernels if we have free high-degree index nodes. 
Therefore, we choose the set $s_i$ that has lower-degree index nodes, leaving high-degree nodes untouched. 
In the second case, if there is no such set covering all kernel nodes, we will choose the set $s_i$ that covers as many kernel nodes as possible, which is aimed to maximize current PEs utilization.
\begin{algorithm}
\caption{Scheduling algorithm}
\label{algo::sheduling}
\begin{algorithmic}[1]
\renewcommand{\algorithmicrequire}{\textbf{Input:}}
\renewcommand{\algorithmicensure}{\textbf{Output:}}
\Require Compressed kernels.
\Ensure Set collection $S^{*}=\left \{s_1, s_2,...,s_i,...\right \}$ for scheduling.
\State Construct bipartite graph $G(KR, ID)$
\While{\text{There are edges existing in} $G(KR, ID)$}
    \State Find set collection $S$ that meet above constraint \textbf{C1}, \textbf{C2}
    \If{\text{there are sets $S^{'}$ covering all kernel nodes}}
        \For{\text{set} $s_i$ \textbf{in} $S^{'}$}
        \State Choose $s_i$ in which edges has minimum index 
        \State node degree in $G(KR, ID)$
        \EndFor
    \Else
        \For{\text{set} $s_i$ \textbf{in} $S$}
        \State Choose $s_i$ which cover most kernel nodes in 
        \State $G(KR, ID)$
        \EndFor
    \EndIf
\State Push selected set $s_i$ into $S^{*}$
\State Delete edges in selected set $s_i$ from $G(KR, ID)$
\EndWhile
\end{algorithmic} 
\end{algorithm}

\begin{figure}[!htb]
    \centering
    \includegraphics[width=0.5\textwidth]{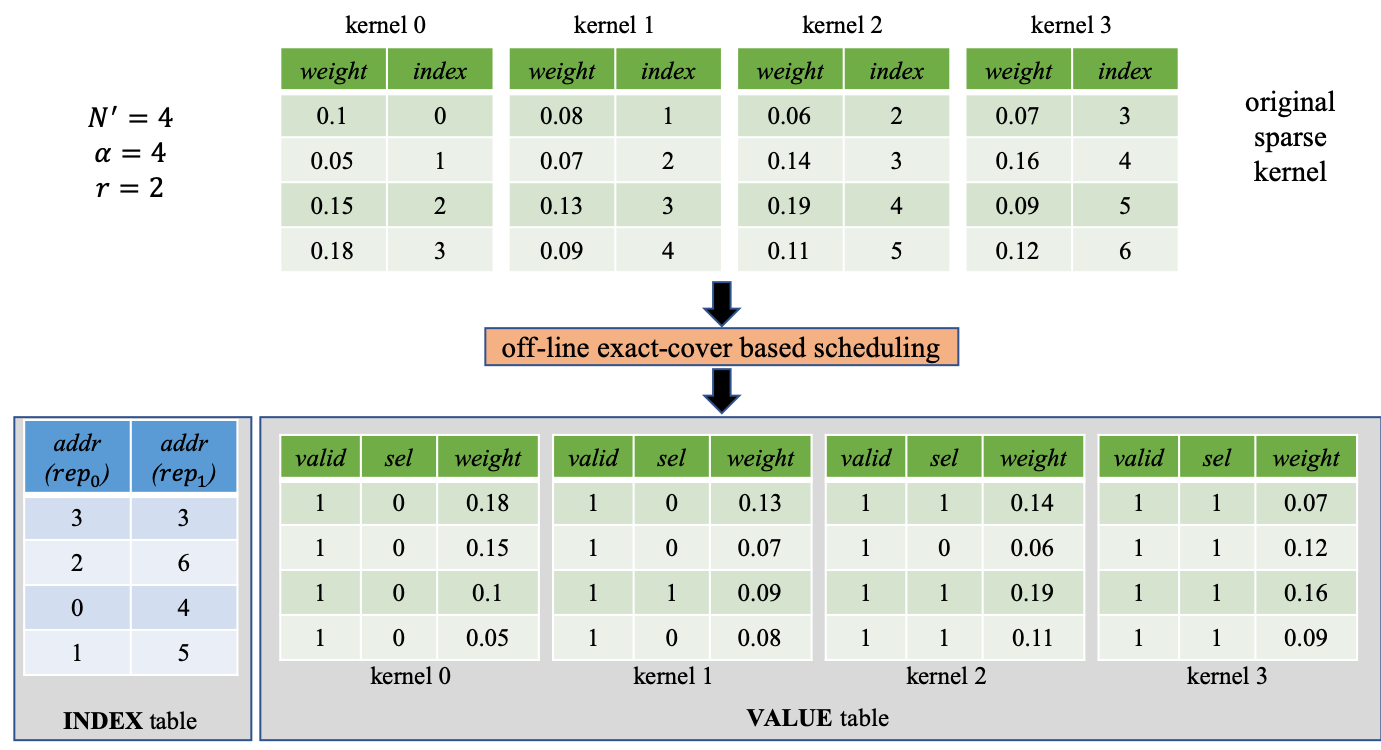}
    \caption{Storage arrangement for sparse kernels}
    \vspace{-.4cm}
    \label{fig::sparsekernmem}
\end{figure}

After getting the scheduling list, to efficiently access values in kernel and replica buffers, we break $S^{*}$ into two parts, \textbf{INDEX} table and \textbf{Value} table. For each $s_i \in S^{*}$, we collect the unique indices, storing them into \textbf{INDEX} table; we store weights and corresponding selection signal $sel$ into \textbf{VALUE} table, as shown in Fig.~\ref{fig::sparsekernmem}. During hardware implementation, we first read these indices ($rep_0$,$rep_1$) and value tables, then we use these indices to read necessary inputs at current cycle, then we use \textit{sel} signal in value tables to route inputs into corresponding PEs. Furthermore, in some cases, some kernels might be inactive due to too many unique addresses in current cycle. Therefore, we use \textit{valid} signal to indicate whether we should feed certain kernels or not.

\section{Design Evaluation}\label{sec::exp}
The fundamental objective in this paper is to reduce communication overhead brought by enlarged spectral kernels and minimize read conflicts when multiple sparse kernels access input tile in one single BRAM. 
In this section, we use VGG16 to show: (1) how our dataflow optimization technique achieves an optimal trade-off between bandwidth and BRAM usage; 
(2) how exact-cover based scheduling method improves PE utilization. 
Sec.~\ref{sec::exp::bw} covers detailed results of BRAM usage and communication bandwidth using various dataflow techniques. 
Then in Sec.~\ref{sec::exp::DSPutil}, we employ kernels with different patterns to show exact-cover based scheduling method delivers higher PE utilization. FPGA implementation results are shown in Sec.~\ref{sec::exp::FPGAimp}.

We use a heterogeneous CPU-FPGA platform, Xilinx Alveo U200, in which operations like OaA\cite{2017_FPGA_CNNOaA}, ReLU, Pooling, fully-connected layers are offloaded to CPU, while FPGA is dedicated to spectral convolutional layers. 16-bit fix-point (FX) number is adopted during computing. RTL code is synthesized and implemented in Xilinx Vivado 2018.3, then integrated into OpenCL environment.

\subsection{Data transfers}\label{sec::exp::bw}
We use the amount of transferred data between on- and off-chip memory as the metric to compare communication overhead in different dataflows. Required bandwidth can be easily obtained given latency $\tau_i$ on each layer.

Table\ref{tab::VGG16flowparam} shows the optimal \textbf{architecture} and \textbf{streaming} parameters for VGG16 given compression ratio $\alpha=4$. Since the first layer (conv1\_1) has negligible computations, we omit it during dataflow optimization.
Given these 
parameters for each layer, and based on previous complexity analysis in Sec.~\ref{sec::analysis} and Sec.~\ref{sec::arch::flexible_flow}, we can compare this optimized flow \textbf{Flow opt} with \textbf{Flow \#1} (streaming kernels) and \textbf{Flow \#2} (streaming input tiles) in terms of BRAM usage and data transfers. 

\begin{figure*}[!htb]
    \centering
    \includegraphics[width=0.8\textwidth]{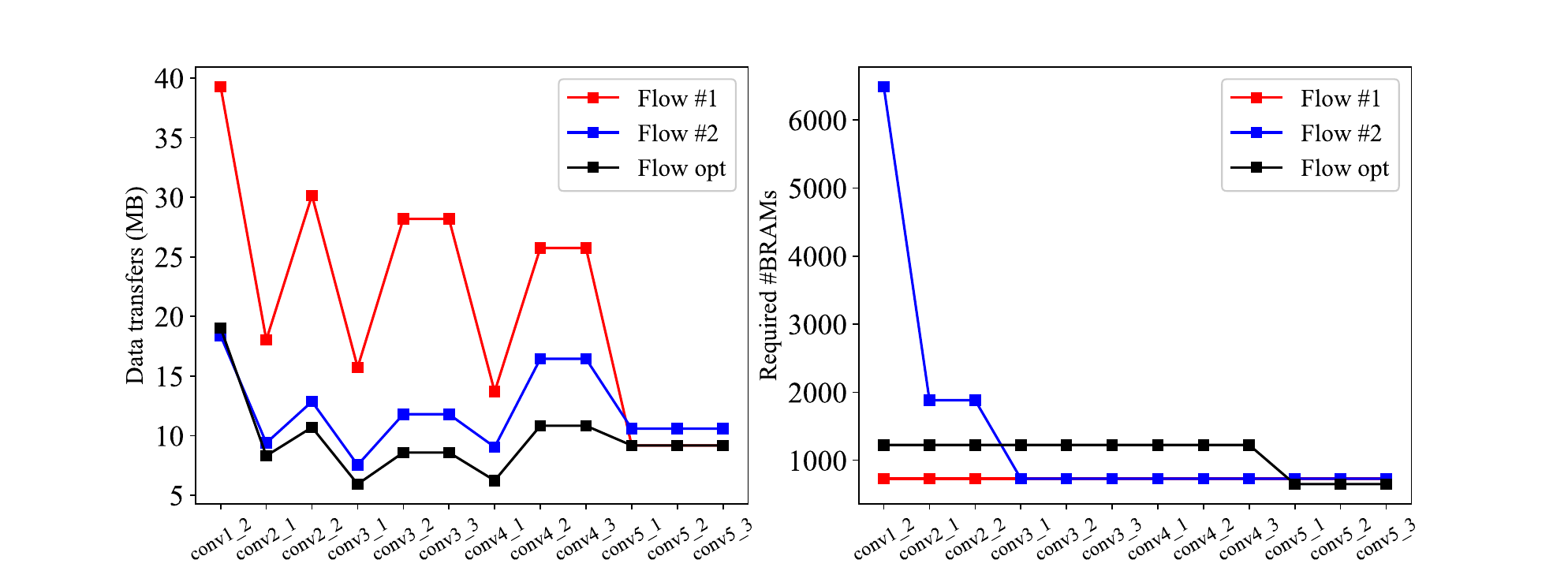}
    \caption{Complexity comparison between different dataflows}
    \label{plt::complexitycompare}
\end{figure*}

Fig.~\ref{plt::complexitycompare} shows complexities of different dataflows for VGG16 with $K=8$ and $\alpha=4$. Compare with \textbf{Flow \#1} and \textbf{Flow \#2}, \textbf{Flow opt} transfers minimal amount of data in almost all layers, while still keeping moderate BRAM usage. 
Though \textbf{Flow opt} consumes more BRAMs than \textbf{Flow \#1} design due to the fact that more partial sums needs to be stored, it significantly reduce times of reloading input tiles or kernels during computing. It accordingly relieves the system of much pressure in transferring data. 
Spectral VGG16 with $K=16$ also delivers similar improvements. However, since the model with $16\times 16$ spectral kernels needs $4\times$ more storage for kernels, even with our optimized dataflow, it still causes huge communication overhead. Hence, during hardware implementation, we will choose $8\times 8$ spectral kernels.
\begin{table}[!htb]
\caption{Architecture and streaming parameter for VGG16}
\label{tab::VGG16flowparam}
\begin{tabular}{cccccc}
\toprule
lyr & conv1\_2 & conv2\_* & conv3\_* & conv4\_* & conv5\_*\\
\midrule
\midrule
\multicolumn{6}{c}{VGG16 ($K=8$, $P^{'}=9$, $N^{'}=64$)} \\
\midrule
$P_s$ & 243 & 126 & 108 & 27 & 9 \\
$N_s$ & 64 & 128 & 128 & 512 & 512 \\
\midrule
\multicolumn{6}{c}{VGG16 ($K=16$, $P^{'}=16$, $N^{'}=32$)} \\
\midrule
$P_s$ & 64 & 32 & 16 & 16 & 16 \\
$N_s$ & 64 & 128 & 256 & 256 & 256 \\
\bottomrule
\end{tabular}
\vspace{-.4cm}
\end{table}

The required bandwidth can be easily calculated given latency budget $\tau_i$ in each layer $i$. 
We first assume the platform has enough bandwidth during computing, then $\tau_i$ is determined based on each layer's computation complexity. 
Suppose the total latency budget for convolutional layers is $\tau$, then for each layer $\tau_i = \tau \times \frac{CMP_i}{CMP_{\text{total}}}$, where $CPM_i$ denotes multiplications and additions in layer $i$. 
Table\ref{tab::VGG16bw} gives the required bandwidth given total inference latency $\tau=20$ ms\footnote{20 ms corresponds to sample rate of $50\sim 60$ fps in video cameras, which is common in real-time image processing.}. 
Even with enlarged spectral kernels, the whole design still reaps significant improvements from \textbf{Flow opt} and avoid being communication-bounded.

\begin{table*}[!htb]
\caption{Required bandwidth for VGG16 under \textbf{Flow opt}}
\label{tab::VGG16bw}
\begin{tabular}{ccccccccccc}
\toprule
\multicolumn{9}{c}{VGG16 ($K=8$, $P^{'}=9$, $N^{'}=64$)}  \\
\midrule
 lyr & conv1\_2 & conv2\_1 & conv2\_2 & conv3\_1 & conv3\_2,3 & conv4\_1 & conv4\_2,3 & conv5\_1,2,3\\
 BW (GB/s)& 8.2 & 7.3 & 4.7 & 4.8 & 3.5 & 5.0 & 4.3 & 9.9 \\
 \toprule
\end{tabular}
\vspace{-.4cm}
\end{table*}

\subsection{PE utilization}\label{sec::exp::DSPutil}
When evaluating the scheduling algorithm, we use PE utilization $\mu_i$ to show the average number active PEs in each layer $i$. Given $r$ replicas for each input image tile, since the total number of multiplications and additions are fixed, higher PE utilization means less number of processing cycles in each convolutional layer.
We first define PE utilization $\mu_i$ in layer $i$ as:
\begin{equation}\label{eq::PEutil}
    \mu_i = \frac{\sum_{t=1}^{T}PE_{\text{on}_t}}{T\cdot PE_{\text{total}}} = \frac{\sum_{t=1}^{T}PE_{\text{on}_t}}{T\cdot N^{'}P^{'}},
\end{equation}
where $T$ denotes the total convolution cycles, each cycle contributed by Hadamard product with $P^{'}$ input tiles and $N^{'}$ kernels; $PE_{\text{on}_t}$ and $PE_{\text{total}}$ are the number of working PEs on cycle $t$, and the total number of PEs on chip.

To compare, we also implement two baseline methods: \textbf{random scheduling} and \textbf{lowest-index first scheduling}\cite{2019_HiPC_SPEC2}. 
\textbf{random scheduling} randomly chooses both a kernel and a non-zero weight index in this kernel, then continues randomly choosing other kernels and indices until either all kernels are included or the number of unique indices reaches $r$. 
On the other hand, \textbf{lowest-index first scheduling} always picks the kernels with lowest index in the current group. It iterates this operation until the same stopping condition is triggered.

\begin{figure}[!htb]
    \centering
    \includegraphics[width=0.45\textwidth]{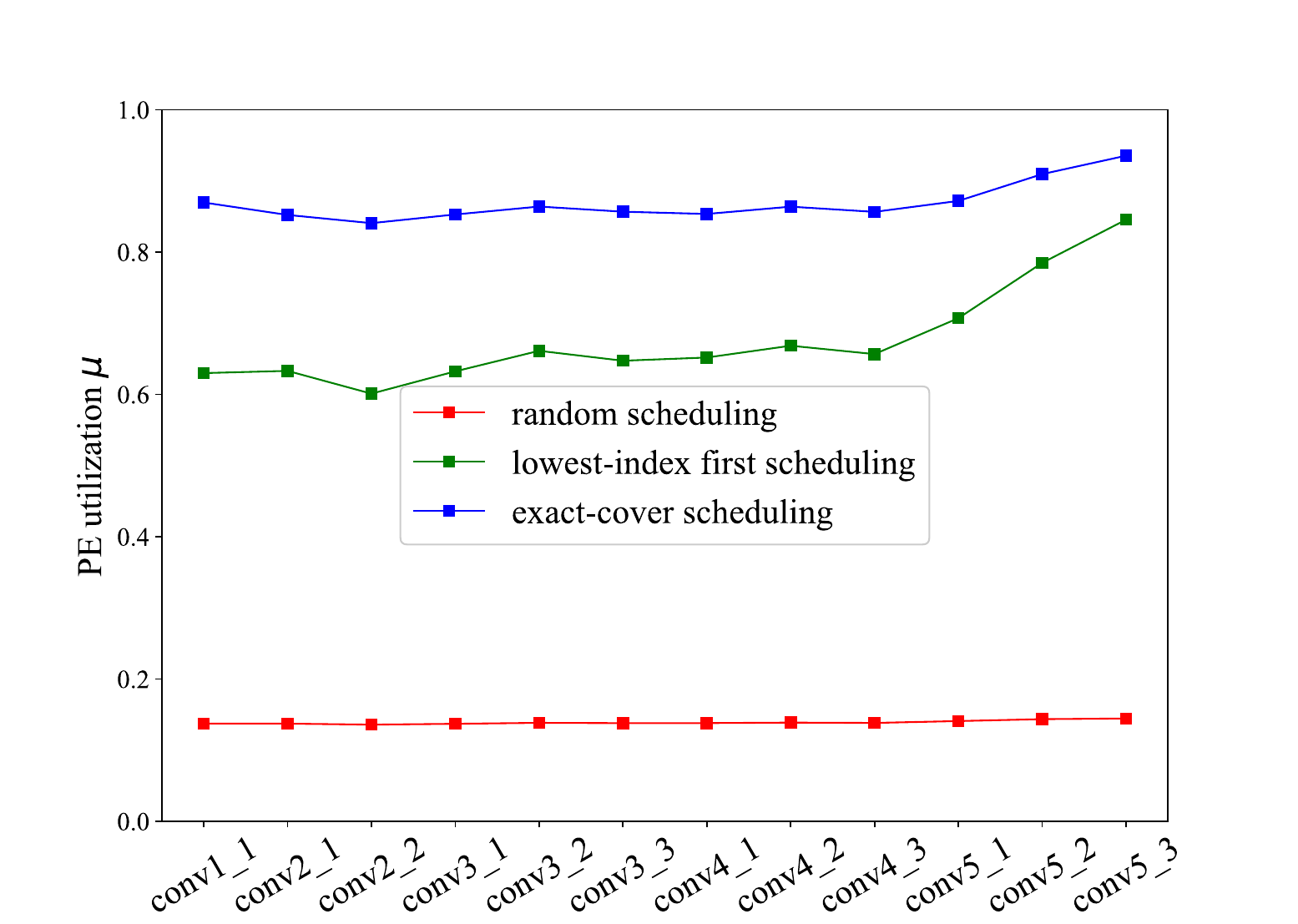}
    \caption{PE utilization given in each layer of VGG16 $r=8$}
    \vspace{-.4cm}
    \label{plt::PEutilVGG16layerwise}
\end{figure}

Fig.\ref{plt::PEutilVGG16layerwise} shows PE utilization in each layer of VGG16 using above three scheduling methods with the number of replicas $r=8$, and $N^{'}=64$. 
Exact-cover based scheduling method achieves the highest PE utilization. 
Moreover, it also gives a consistent performance across all convolutional layers, while lowest-index first scheduling deeply relies on the condition that indices in different kernels are close, like kernels in layer conv5\_2, conv5\_3. 

\begin{figure}[!htb]
    \centering
    \includegraphics[width=0.4\textwidth]{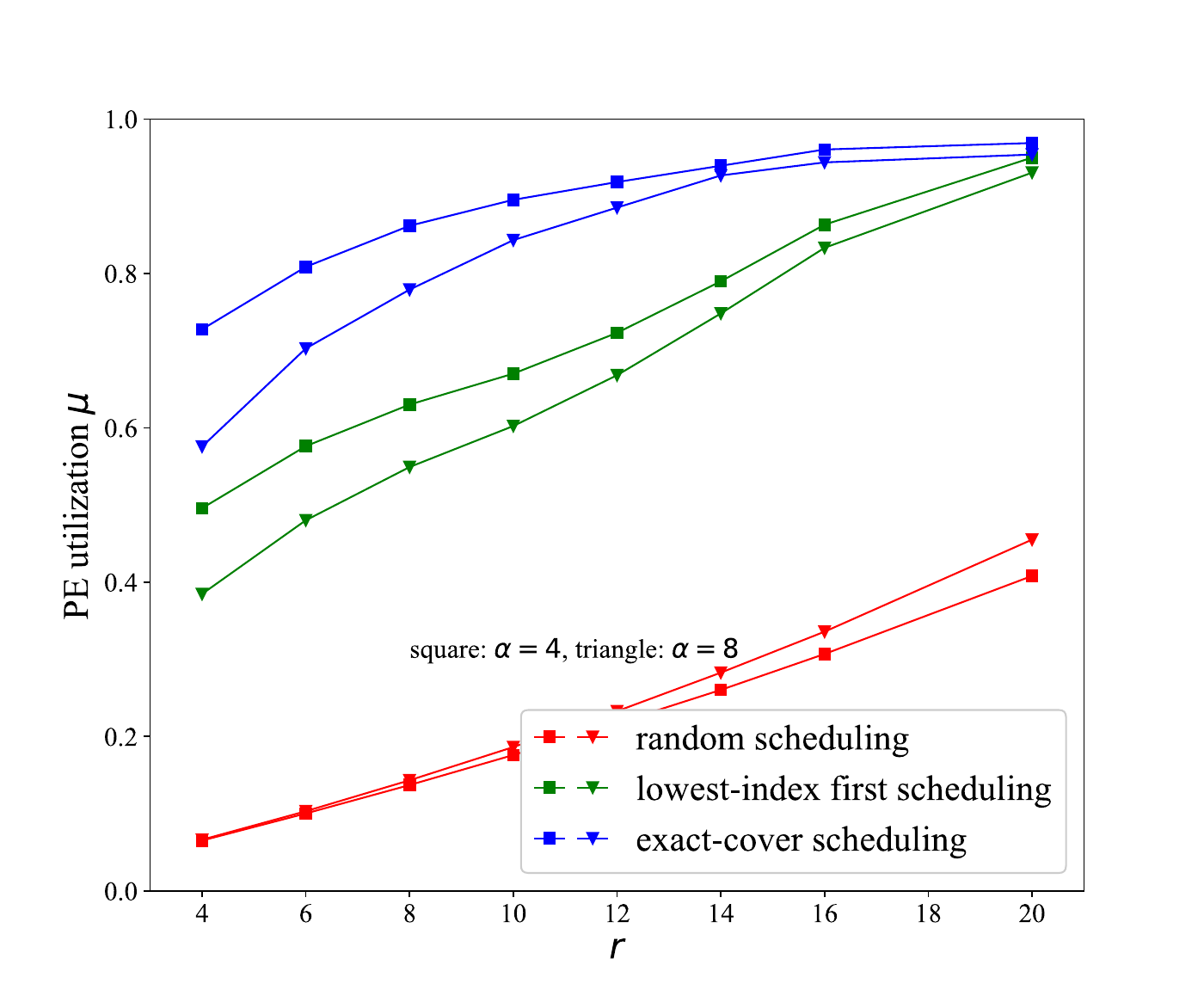}
    \caption{Average PE utilization for VGG16 given different number of replicas}
    \label{plt::PEutilVGG16replwise}
    \vspace{-.1cm}
\end{figure}

To show how PE utilization changes with the number of replicas $r$ under different scheduling methods, we vary the number of replicas $r$ from 4 to 20. 
Fig.\ref{plt::PEutilVGG16replwise} shows the average PE utilization under different compression ratios, in which the average PE utilization is obtained by weighting each layer's utilization with its total computations. 
Scheduling method in this paper reaches very high PE utilization ($> 80\%$) given fewer number of replicas compared with lowest-index first method. 
Even under $\times 8$ compression, in which indices are largely scattered from 0 to $K^2-1$, exact-cover based scheduling still achieves $> 80\%$ PE utilization given 10 replicas for each input tiles.
Lowest-index first method, on the contrary, needs 16 replicas to reach comparable performance. 

\begin{figure}[!htb]
    \centering
    \includegraphics[width=0.4\textwidth]{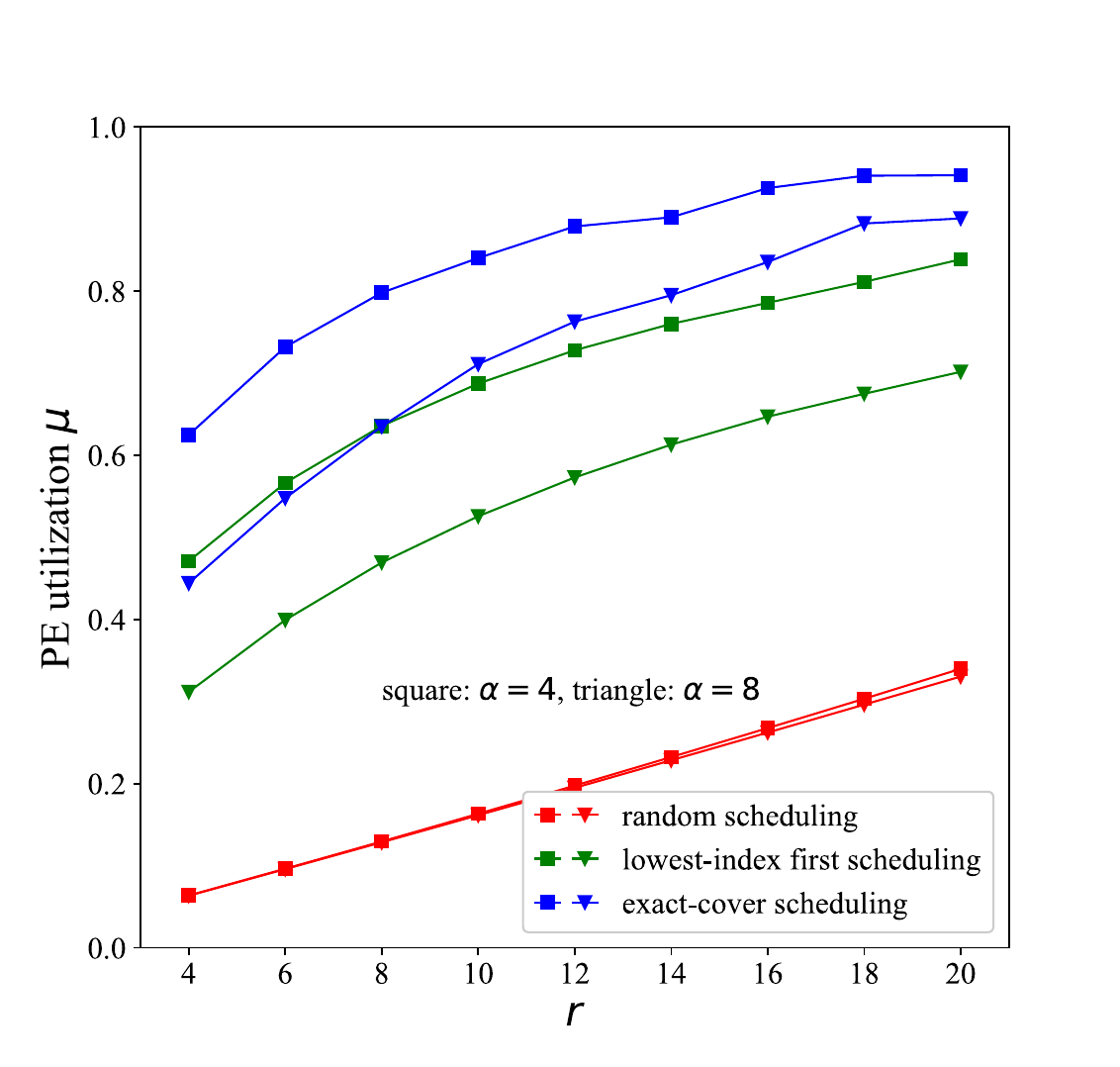}
    \caption{Average PE utilization for random non-zeros given different number of replicas}
    \vspace{-.4cm}
    \label{plt::PEutilVGG16replwiserandom}
\end{figure}

Apart from sparse kernels generated by algorithm in \cite{2019_HiPC_SPEC2}, robustness to more general sparsity patterns is also critical. We generate sparse kernels from uncompressed spectral model by randomly choose $\frac{K^2}{\alpha}$ non-zero weights while zeroing other values. 
We try various compression ratios, as well as different number of replicas $r$. Fig.~\ref{plt::PEutilVGG16replwiserandom} shows the average PE utilization for these three scheduling methods. 
Exact-cover based methods always outperforms lowest-index first scheduling.
In addition, even with random sparsity pattern, exact-cover based methods still achieves comparable PE utilization as in ADMM-based pruned kernels when $\alpha=4$.

\begin{table*}[t]
\caption{Implementation on FPGA and comparison with other works}
\label{tab::hardwareimpl}
\begin{tabular}{cccccc}
\toprule
 & \cite{2017_FPGA_CNNOaA} & \cite{2018_FPGA_CNNCaP} & \cite{2019_HiPC_SPEC2} & \cite{2017_JETC_Sparcnet} & This work \\
 \midrule
 \midrule
Device & Intel QPI FPGA & Stratix V & Virtex XC7VX690T & Artix 7 XC7A200T & Xilinx Alveo U200 \\
Datatype & 16bit FT & 16bit FX & 16bit FX & 16bit FX & 16bit FX \\
DSP & 224/224 & 256/256 & 3200/3600 & 384/740 & 2680/6840 \\
BRAM & - & 1377/1880 & 1244/1470 & 194/365 & 1469/2160 \\
LUT & - & 107K/233K & 237K/430K & - & 230K/1.2M \\ 
Clock(MHz) & 200 & 200 & 200 & 100 & 200 \\ 
\midrule
Throughput(fps) & 4 & 6 & 148 & 5 & 112 \\
Latency(ms) & 250 & 167 & 68 & 200 & 9\\
Bandwidth(GB/s) & 5.0 & - & 9 & - & 12\\
\bottomrule
\end{tabular}
\end{table*}

\subsection{FPGA implementation}\label{sec::exp::FPGAimp}
During implementation, we choose VGG16 as our target model fed with $224\times 224$ images, in which compression ratio $\alpha=4$, corresponding accuracy is $95.0\%$. 
Architecture parameters $P^{'}, N^{'}$ are set to be 9 and 64, while streaming parameters in each layer are the same as Table\ref{tab::VGG16flowparam}. 
Based on the scheduling analysis, we choose the number of replicas $r$ to be 10 to get $> 90\%$ PE utilization. 
We use the state-of-the-art FPGA, Xilinx Alveo U200, as our target platform, which has $6840$ DSPs and $2160$ RBAMs. 
URAM (288Kb each) is introduced recently as a global on-chip memory, which we didn't consider in our analysis. However, our dataflow analysis can be still generalized to this new memory, for which we can translate off-chip data transfers as communication between global buffers and local buffers.
Table\ref{tab::hardwareimpl} shows the implementation result of our whole architecture, as well as performance comparison with other works. 
With using $40\%$ DSPs and $68\%$ BRAMs, we still achieve 200MHz clock frequency. Given $N^{'}=64$ and $P^{'}=9$, the inference in all convolutional layers  is done within $9$ ms, while the required bandwidth is lowered to $12$ GB/s. 
\cite{2017_FPGA_CNNOaA} and \cite{2018_FPGA_CNNCaP} are aimed to accelerate uncompressed spectral CNNs on FPGAs. In terms of latency, we reduce the inference time by $46\%$ after simply scaling the resources utilization.
Compared with \cite{2019_HiPC_SPEC2}, though it has higher throughput, it can only process images in batches. Single-image inference latency is 68ms, which is $7.5\times$ higher than the design in this paper. If we scale its latency to 9ms, the required bandwidth explodes to almost 70GB/s, which is far beyond single DDR's capacity. 
\cite{2017_JETC_Sparcnet} is a design for sparse spatial CNNs. Since it uses an old device, to give a fair comparison, we also assume it can be deployed in Alveo U200, while accessing the same resources. Under 200MHz, we can still get near $40\%$ latency improvement.

\begin{figure}[!htb]
    \centering
    \includegraphics[width=0.45\textwidth]{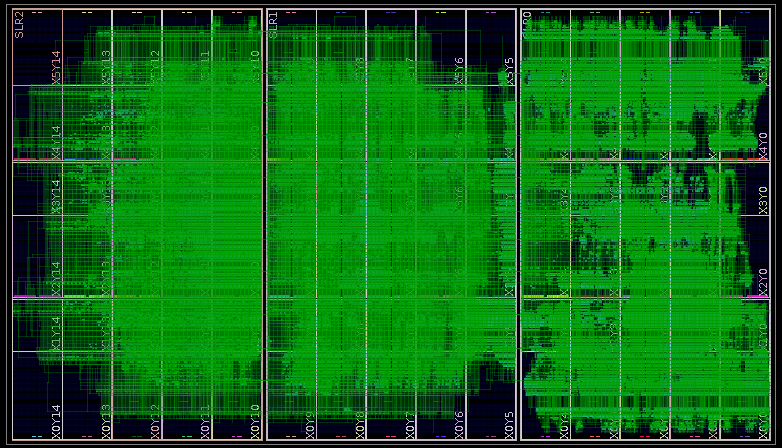}
    \caption{FPGA footprint of our design}
    \vspace{-.4cm}
    \label{plt::FPGAfootprint}
\end{figure}

To intuitively show resource utilization in the FPGA device, we give a FPGA footprint after implementation, as shown in Fig.~\ref{plt::FPGAfootprint}. Our design take up almost $70\%$ of the total area and most routing resources, which explains why we don't further increase BRAM and DSP usage.

\section{Conclusion}
Pruned spectral CNNs can reduce computation complexity without accuracy loss but have a kernel storage explosion problem. We develop a principled design approach to fully exploit the potential of sparse spectral CNNs and overcome the performance gap between pruning and FPGA implementation: first, we analyze the bandwidth-storage tradeoff of sparse convolutional layers and locate bottlenecks. We then adopt a flexible dataflow for different layers to minimize communication overhead. To ameliorate the performance degradation brought on by irregular on-chip memory accesses of sparse kernels, we design an approximate exact-cover based scheduling algorithm to efficiently feed PEs with few replicas. Finally, we implement a unified hardware architecture which can adjust flexible dataflow on the fly and maximize PE utilization. On a Xilinx Alveo U200 platform, with 64 parallel kernels and 9 parallel input tiles, inference takes 9ms, while only requiring 12GB/s off-chip \looseness=-1
bandwidth.

\section{Acknowledgements}
This work was supported in part by the National Science Foundation under grants CNS-1643351 and CCF-1919117.

\newpage

\bibliographystyle{ACM-Reference-Format}
\balance
\bibliography{sample-base}


\begin{thebibliography}{30}


\ifx \showCODEN    \undefined \def \showCODEN     #1{\unskip}     \fi
\ifx \showDOI      \undefined \def \showDOI       #1{#1}\fi
\ifx \showISBNx    \undefined \def \showISBNx     #1{\unskip}     \fi
\ifx \showISBNxiii \undefined \def \showISBNxiii  #1{\unskip}     \fi
\ifx \showISSN     \undefined \def \showISSN      #1{\unskip}     \fi
\ifx \showLCCN     \undefined \def \showLCCN      #1{\unskip}     \fi
\ifx \shownote     \undefined \def \shownote      #1{#1}          \fi
\ifx \showarticletitle \undefined \def \showarticletitle #1{#1}   \fi
\ifx \showURL      \undefined \def \showURL       {\relax}        \fi
\providecommand\bibfield[2]{#2}
\providecommand\bibinfo[2]{#2}
\providecommand\natexlab[1]{#1}
\providecommand\showeprint[2][]{arXiv:#2}

\bibitem[\protect\citeauthoryear{Amos, Ludwiczuk, Satyanarayanan,
  et~al\mbox{.}}{Amos et~al\mbox{.}}{2016}]%
        {2016_CMU_Openface}
\bibfield{author}{\bibinfo{person}{Brandon Amos}, \bibinfo{person}{Bartosz
  Ludwiczuk}, \bibinfo{person}{Mahadev Satyanarayanan}, {et~al\mbox{.}}}
  \bibinfo{year}{2016}\natexlab{}.
\newblock \showarticletitle{Openface: A general-purpose face recognition
  library with mobile applications}.
\newblock \bibinfo{journal}{\emph{CMU School of Computer Science}}
  \bibinfo{volume}{6} (\bibinfo{year}{2016}).
\newblock


\bibitem[\protect\citeauthoryear{Boyd, Parikh, Chu, Peleato, and Eckstein}{Boyd
  et~al\mbox{.}}{2011}]%
        {admm:boyd}
\bibfield{author}{\bibinfo{person}{Stephen Boyd}, \bibinfo{person}{Neal
  Parikh}, \bibinfo{person}{Eric Chu}, \bibinfo{person}{Borja Peleato}, {and}
  \bibinfo{person}{Jonathan Eckstein}.} \bibinfo{year}{2011}\natexlab{}.
\newblock \showarticletitle{Distributed Optimization and Statistical Learning
  via the Alternating Direction Method of Multipliers}.
\newblock \bibinfo{journal}{\emph{Found. Trends Mach. Learn.}}
  \bibinfo{volume}{3}, \bibinfo{number}{1} (\bibinfo{date}{Jan.}
  \bibinfo{year}{2011}), \bibinfo{pages}{1--122}.
\newblock
\showISSN{1935-8237}
\urldef\tempurl%
\url{https://doi.org/10.1561/2200000016}
\showDOI{\tempurl}


\bibitem[\protect\citeauthoryear{Chen, Hu, Wang, Zhao, Amos, Wu, Ha, Elgazzar,
  Pillai, Klatzky, et~al\mbox{.}}{Chen et~al\mbox{.}}{2017}]%
        {2017_SEC_empericalLatencyStudy}
\bibfield{author}{\bibinfo{person}{Zhuo Chen}, \bibinfo{person}{Wenlu Hu},
  \bibinfo{person}{Junjue Wang}, \bibinfo{person}{Siyan Zhao},
  \bibinfo{person}{Brandon Amos}, \bibinfo{person}{Guanhang Wu},
  \bibinfo{person}{Kiryong Ha}, \bibinfo{person}{Khalid Elgazzar},
  \bibinfo{person}{Padmanabhan Pillai}, \bibinfo{person}{Roberta Klatzky},
  {et~al\mbox{.}}} \bibinfo{year}{2017}\natexlab{}.
\newblock \showarticletitle{An empirical study of latency in an emerging class
  of edge computing applications for wearable cognitive assistance}. In
  \bibinfo{booktitle}{\emph{Proceedings of the Second ACM/IEEE Symposium on
  Edge Computing}}. ACM, \bibinfo{pages}{14}.
\newblock


\bibitem[\protect\citeauthoryear{Choquette, Giroux, and Foley}{Choquette
  et~al\mbox{.}}{2018}]%
        {2018_MICRO_Volta}
\bibfield{author}{\bibinfo{person}{Jack Choquette}, \bibinfo{person}{Olivier
  Giroux}, {and} \bibinfo{person}{Denis Foley}.}
  \bibinfo{year}{2018}\natexlab{}.
\newblock \showarticletitle{Volta: Performance and programmability}.
\newblock \bibinfo{journal}{\emph{IEEE Micro}} \bibinfo{volume}{38},
  \bibinfo{number}{2} (\bibinfo{year}{2018}), \bibinfo{pages}{42--52}.
\newblock


\bibitem[\protect\citeauthoryear{Courbariaux, Bengio, and David}{Courbariaux
  et~al\mbox{.}}{2015}]%
        {2015_NIPS_BinaryConnect}
\bibfield{author}{\bibinfo{person}{Matthieu Courbariaux},
  \bibinfo{person}{Yoshua Bengio}, {and} \bibinfo{person}{Jean-Pierre David}.}
  \bibinfo{year}{2015}\natexlab{}.
\newblock \showarticletitle{Binaryconnect: Training deep neural networks with
  binary weights during propagations}. In \bibinfo{booktitle}{\emph{Advances in
  neural information processing systems}}. \bibinfo{pages}{3123--3131}.
\newblock


\bibitem[\protect\citeauthoryear{Gupta, Agrawal, Gopalakrishnan, and
  Narayanan}{Gupta et~al\mbox{.}}{2015}]%
        {2015_ICML_LimitedPrecision}
\bibfield{author}{\bibinfo{person}{Suyog Gupta}, \bibinfo{person}{Ankur
  Agrawal}, \bibinfo{person}{Kailash Gopalakrishnan}, {and}
  \bibinfo{person}{Pritish Narayanan}.} \bibinfo{year}{2015}\natexlab{}.
\newblock \showarticletitle{Deep learning with limited numerical precision}. In
  \bibinfo{booktitle}{\emph{International Conference on Machine Learning}}.
  \bibinfo{pages}{1737--1746}.
\newblock


\bibitem[\protect\citeauthoryear{Han, Mao, and Dally}{Han
  et~al\mbox{.}}{2015}]%
        {2015_arXiv_DeepComp}
\bibfield{author}{\bibinfo{person}{Song Han}, \bibinfo{person}{Huizi Mao},
  {and} \bibinfo{person}{William~J Dally}.} \bibinfo{year}{2015}\natexlab{}.
\newblock \showarticletitle{Deep compression: Compressing deep neural networks
  with pruning, trained quantization and huffman coding}.
\newblock \bibinfo{journal}{\emph{arXiv preprint arXiv:1510.00149}}
  (\bibinfo{year}{2015}).
\newblock


\bibitem[\protect\citeauthoryear{He, Zhang, Ren, and Sun}{He
  et~al\mbox{.}}{2016}]%
        {2016_CVPR_ResNet}
\bibfield{author}{\bibinfo{person}{Kaiming He}, \bibinfo{person}{Xiangyu
  Zhang}, \bibinfo{person}{Shaoqing Ren}, {and} \bibinfo{person}{Jian Sun}.}
  \bibinfo{year}{2016}\natexlab{}.
\newblock \showarticletitle{Deep residual learning for image recognition}. In
  \bibinfo{booktitle}{\emph{Proceedings of the IEEE conference on computer
  vision and pattern recognition}}. \bibinfo{pages}{770--778}.
\newblock


\bibitem[\protect\citeauthoryear{He, Zhang, and Sun}{He et~al\mbox{.}}{2017}]%
        {2017_ICCV_ChannelPruning}
\bibfield{author}{\bibinfo{person}{Yihui He}, \bibinfo{person}{Xiangyu Zhang},
  {and} \bibinfo{person}{Jian Sun}.} \bibinfo{year}{2017}\natexlab{}.
\newblock \showarticletitle{Channel pruning for accelerating very deep neural
  networks}. In \bibinfo{booktitle}{\emph{Proceedings of the IEEE International
  Conference on Computer Vision}}. \bibinfo{pages}{1389--1397}.
\newblock


\bibitem[\protect\citeauthoryear{Jacob, Kligys, Chen, Zhu, Tang, Howard, Adam,
  and Kalenichenko}{Jacob et~al\mbox{.}}{2018}]%
        {2018_CVPR_IntegerOnly}
\bibfield{author}{\bibinfo{person}{Benoit Jacob}, \bibinfo{person}{Skirmantas
  Kligys}, \bibinfo{person}{Bo Chen}, \bibinfo{person}{Menglong Zhu},
  \bibinfo{person}{Matthew Tang}, \bibinfo{person}{Andrew Howard},
  \bibinfo{person}{Hartwig Adam}, {and} \bibinfo{person}{Dmitry Kalenichenko}.}
  \bibinfo{year}{2018}\natexlab{}.
\newblock \showarticletitle{Quantization and training of neural networks for
  efficient integer-arithmetic-only inference}. In
  \bibinfo{booktitle}{\emph{Proceedings of the IEEE Conference on Computer
  Vision and Pattern Recognition}}. \bibinfo{pages}{2704--2713}.
\newblock


\bibitem[\protect\citeauthoryear{Krizhevsky, Sutskever, and Hinton}{Krizhevsky
  et~al\mbox{.}}{2012}]%
        {2012_NIPS_AlexNet}
\bibfield{author}{\bibinfo{person}{Alex Krizhevsky}, \bibinfo{person}{Ilya
  Sutskever}, {and} \bibinfo{person}{Geoffrey~E Hinton}.}
  \bibinfo{year}{2012}\natexlab{}.
\newblock \showarticletitle{Imagenet classification with deep convolutional
  neural networks}. In \bibinfo{booktitle}{\emph{Advances in neural information
  processing systems}}. \bibinfo{pages}{1097--1105}.
\newblock


\bibitem[\protect\citeauthoryear{Lin, Talathi, and Annapureddy}{Lin
  et~al\mbox{.}}{2016}]%
        {2016_ICML_FixpointQuant}
\bibfield{author}{\bibinfo{person}{Darryl Lin}, \bibinfo{person}{Sachin
  Talathi}, {and} \bibinfo{person}{Sreekanth Annapureddy}.}
  \bibinfo{year}{2016}\natexlab{}.
\newblock \showarticletitle{Fixed point quantization of deep convolutional
  networks}. In \bibinfo{booktitle}{\emph{International Conference on Machine
  Learning}}. \bibinfo{pages}{2849--2858}.
\newblock


\bibitem[\protect\citeauthoryear{Long, Shelhamer, and Darrell}{Long
  et~al\mbox{.}}{2015}]%
        {2015_FCNN_CVPR}
\bibfield{author}{\bibinfo{person}{Jonathan Long}, \bibinfo{person}{Evan
  Shelhamer}, {and} \bibinfo{person}{Trevor Darrell}.}
  \bibinfo{year}{2015}\natexlab{}.
\newblock \showarticletitle{Fully convolutional networks for semantic
  segmentation}. In \bibinfo{booktitle}{\emph{Proceedings of the IEEE
  conference on computer vision and pattern recognition}}.
  \bibinfo{pages}{3431--3440}.
\newblock


\bibitem[\protect\citeauthoryear{Luo, Wu, and Lin}{Luo et~al\mbox{.}}{2017}]%
        {2017_ICCV_ThiNet}
\bibfield{author}{\bibinfo{person}{Jian-Hao Luo}, \bibinfo{person}{Jianxin Wu},
  {and} \bibinfo{person}{Weiyao Lin}.} \bibinfo{year}{2017}\natexlab{}.
\newblock \showarticletitle{ThiNet: A Filter Level Pruning Method for Deep
  Neural Network Compression}. In \bibinfo{booktitle}{\emph{The IEEE
  International Conference on Computer Vision (ICCV)}}.
\newblock


\bibitem[\protect\citeauthoryear{Mei, Liu, Niu, Ji, Zhou, and Wang}{Mei
  et~al\mbox{.}}{2017}]%
        {Lowrank_GlobalSIP_2017}
\bibfield{author}{\bibinfo{person}{Chunsheng Mei}, \bibinfo{person}{Zhenyu
  Liu}, \bibinfo{person}{Yue Niu}, \bibinfo{person}{Xiangyang Ji},
  \bibinfo{person}{Wei Zhou}, {and} \bibinfo{person}{Dongsheng Wang}.}
  \bibinfo{year}{2017}\natexlab{}.
\newblock \showarticletitle{A 200mhz 202.4 gflops@ 10.8 w vgg16 accelerator in
  xilinx vx690t}. In \bibinfo{booktitle}{\emph{2017 IEEE Global Conference on
  Signal and Information Processing (GlobalSIP)}}. IEEE,
  \bibinfo{pages}{784--788}.
\newblock


\bibitem[\protect\citeauthoryear{Niu, Zeng, Srivastava, Lakhotia, Kannan, Wang,
  and Prasanna}{Niu et~al\mbox{.}}{2019}]%
        {2019_HiPC_SPEC2}
\bibfield{author}{\bibinfo{person}{Yue Niu}, \bibinfo{person}{Hanqing Zeng},
  \bibinfo{person}{Ajitesh Srivastava}, \bibinfo{person}{Kartik Lakhotia},
  \bibinfo{person}{Rajgopal Kannan}, \bibinfo{person}{Yanzhi Wang}, {and}
  \bibinfo{person}{Viktor Prasanna}.} \bibinfo{year}{2019}\natexlab{}.
\newblock \showarticletitle{SPEC2: \underline{SPEC}tral
  \underline{SP}ars\underline{E} \underline{C}NN Accelerator on FPGAs}. In
  \bibinfo{booktitle}{\emph{Proceedings of the 2019 IEEE International
  Conference on High Performance Computing, Data, and Analytics}}. IEEE.
\newblock


\bibitem[\protect\citeauthoryear{Page, Jafari, Shea, and Mohsenin}{Page
  et~al\mbox{.}}{2017}]%
        {2017_JETC_Sparcnet}
\bibfield{author}{\bibinfo{person}{Adam Page}, \bibinfo{person}{Ali Jafari},
  \bibinfo{person}{Colin Shea}, {and} \bibinfo{person}{Tinoosh Mohsenin}.}
  \bibinfo{year}{2017}\natexlab{}.
\newblock \showarticletitle{Sparcnet: A hardware accelerator for efficient
  deployment of sparse convolutional networks}.
\newblock \bibinfo{journal}{\emph{ACM Journal on Emerging Technologies in
  Computing Systems (JETC)}} (\bibinfo{year}{2017}).
\newblock


\bibitem[\protect\citeauthoryear{Qiu, Wang, Yao, Guo, Li, Zhou, Yu, Tang, Xu,
  Song, et~al\mbox{.}}{Qiu et~al\mbox{.}}{2016}]%
        {2016_FPGA_SVD}
\bibfield{author}{\bibinfo{person}{Jiantao Qiu}, \bibinfo{person}{Jie Wang},
  \bibinfo{person}{Song Yao}, \bibinfo{person}{Kaiyuan Guo},
  \bibinfo{person}{Boxun Li}, \bibinfo{person}{Erjin Zhou},
  \bibinfo{person}{Jincheng Yu}, \bibinfo{person}{Tianqi Tang},
  \bibinfo{person}{Ningyi Xu}, \bibinfo{person}{Sen Song}, {et~al\mbox{.}}}
  \bibinfo{year}{2016}\natexlab{}.
\newblock \showarticletitle{Going deeper with embedded fpga platform for
  convolutional neural network}. In \bibinfo{booktitle}{\emph{Proceedings of
  the 2016 ACM/SIGDA International Symposium on Field-Programmable Gate
  Arrays}}. ACM, \bibinfo{pages}{26--35}.
\newblock


\bibitem[\protect\citeauthoryear{Ren, He, Girshick, and Sun}{Ren
  et~al\mbox{.}}{2015}]%
        {2015_NIPS_FasterRNN}
\bibfield{author}{\bibinfo{person}{Shaoqing Ren}, \bibinfo{person}{Kaiming He},
  \bibinfo{person}{Ross Girshick}, {and} \bibinfo{person}{Jian Sun}.}
  \bibinfo{year}{2015}\natexlab{}.
\newblock \showarticletitle{Faster r-cnn: Towards real-time object detection
  with region proposal networks}. In \bibinfo{booktitle}{\emph{Advances in
  neural information processing systems}}. \bibinfo{pages}{91--99}.
\newblock


\bibitem[\protect\citeauthoryear{Simonyan and Zisserman}{Simonyan and
  Zisserman}{2014}]%
        {2014_arXiv_VGG}
\bibfield{author}{\bibinfo{person}{Karen Simonyan} {and}
  \bibinfo{person}{Andrew Zisserman}.} \bibinfo{year}{2014}\natexlab{}.
\newblock \showarticletitle{Very deep convolutional networks for large-scale
  image recognition}.
\newblock \bibinfo{journal}{\emph{arXiv preprint arXiv:1409.1556}}
  (\bibinfo{year}{2014}).
\newblock


\bibitem[\protect\citeauthoryear{Sodani, Gramunt, Corbal, Kim, Vinod,
  Chinthamani, Hutsell, Agarwal, and Liu}{Sodani et~al\mbox{.}}{2016}]%
        {2016_MICRO_Knights}
\bibfield{author}{\bibinfo{person}{Avinash Sodani}, \bibinfo{person}{Roger
  Gramunt}, \bibinfo{person}{Jesus Corbal}, \bibinfo{person}{Ho-Seop Kim},
  \bibinfo{person}{Krishna Vinod}, \bibinfo{person}{Sundaram Chinthamani},
  \bibinfo{person}{Steven Hutsell}, \bibinfo{person}{Rajat Agarwal}, {and}
  \bibinfo{person}{Yen-Chen Liu}.} \bibinfo{year}{2016}\natexlab{}.
\newblock \showarticletitle{Knights landing: Second-generation intel xeon phi
  product}.
\newblock \bibinfo{journal}{\emph{Ieee micro}} \bibinfo{volume}{36},
  \bibinfo{number}{2} (\bibinfo{year}{2016}), \bibinfo{pages}{34--46}.
\newblock


\bibitem[\protect\citeauthoryear{Szegedy, Liu, Jia, Sermanet, Reed, Anguelov,
  Erhan, Vanhoucke, and Rabinovich}{Szegedy et~al\mbox{.}}{2015}]%
        {2015_CVPR_GoogLeNet}
\bibfield{author}{\bibinfo{person}{Christian Szegedy}, \bibinfo{person}{Wei
  Liu}, \bibinfo{person}{Yangqing Jia}, \bibinfo{person}{Pierre Sermanet},
  \bibinfo{person}{Scott Reed}, \bibinfo{person}{Dragomir Anguelov},
  \bibinfo{person}{Dumitru Erhan}, \bibinfo{person}{Vincent Vanhoucke}, {and}
  \bibinfo{person}{Andrew Rabinovich}.} \bibinfo{year}{2015}\natexlab{}.
\newblock \showarticletitle{Going deeper with convolutions}. In
  \bibinfo{booktitle}{\emph{Proceedings of the IEEE conference on computer
  vision and pattern recognition}}. \bibinfo{pages}{1--9}.
\newblock


\bibitem[\protect\citeauthoryear{Wei, Yu, Zhang, Chen, Wang, Hu, Liang, and
  Cong}{Wei et~al\mbox{.}}{2017}]%
        {2017_DAC_CNNSystolic}
\bibfield{author}{\bibinfo{person}{Xuechao Wei}, \bibinfo{person}{Cody~Hao Yu},
  \bibinfo{person}{Peng Zhang}, \bibinfo{person}{Youxiang Chen},
  \bibinfo{person}{Yuxin Wang}, \bibinfo{person}{Han Hu}, \bibinfo{person}{Yun
  Liang}, {and} \bibinfo{person}{Jason Cong}.} \bibinfo{year}{2017}\natexlab{}.
\newblock \showarticletitle{Automated systolic array architecture synthesis for
  high throughput CNN inference on FPGAs}. In
  \bibinfo{booktitle}{\emph{Proceedings of the 54th Annual Design Automation
  Conference 2017}}. ACM, \bibinfo{pages}{29}.
\newblock


\bibitem[\protect\citeauthoryear{Yaqoob, Khan, Kazmi, Imran, Guizani, and
  Hong}{Yaqoob et~al\mbox{.}}{2019}]%
        {2019_Network_Selfdriving}
\bibfield{author}{\bibinfo{person}{Ibrar Yaqoob}, \bibinfo{person}{Latif~U
  Khan}, \bibinfo{person}{SM~Ahsan Kazmi}, \bibinfo{person}{Muhammad Imran},
  \bibinfo{person}{Nadra Guizani}, {and} \bibinfo{person}{Choong~Seon Hong}.}
  \bibinfo{year}{2019}\natexlab{}.
\newblock \showarticletitle{Autonomous Driving Cars in Smart Cities: Recent
  Advances, Requirements, and Challenges}.
\newblock \bibinfo{journal}{\emph{IEEE Network}} (\bibinfo{year}{2019}).
\newblock


\bibitem[\protect\citeauthoryear{Zeiler and Fergus}{Zeiler and Fergus}{2014}]%
        {2014_ECCV_UnderstandCNN}
\bibfield{author}{\bibinfo{person}{Matthew~D Zeiler} {and} \bibinfo{person}{Rob
  Fergus}.} \bibinfo{year}{2014}\natexlab{}.
\newblock \showarticletitle{Visualizing and understanding convolutional
  networks}. In \bibinfo{booktitle}{\emph{European conference on computer
  vision}}. Springer, \bibinfo{pages}{818--833}.
\newblock


\bibitem[\protect\citeauthoryear{Zeng, Chen, Zhang, and Prasanna}{Zeng
  et~al\mbox{.}}{2018}]%
        {2018_FPGA_CNNCaP}
\bibfield{author}{\bibinfo{person}{Hanqing Zeng}, \bibinfo{person}{Ren Chen},
  \bibinfo{person}{Chi Zhang}, {and} \bibinfo{person}{Viktor Prasanna}.}
  \bibinfo{year}{2018}\natexlab{}.
\newblock \showarticletitle{A framework for generating high throughput CNN
  implementations on FPGAs}. In \bibinfo{booktitle}{\emph{Proceedings of the
  2018 ACM/SIGDA International Symposium on Field-Programmable Gate Arrays}}.
  ACM, \bibinfo{pages}{117--126}.
\newblock


\bibitem[\protect\citeauthoryear{Zhang and Prasanna}{Zhang and
  Prasanna}{2017}]%
        {2017_FPGA_CNNOaA}
\bibfield{author}{\bibinfo{person}{Chi Zhang} {and} \bibinfo{person}{Viktor
  Prasanna}.} \bibinfo{year}{2017}\natexlab{}.
\newblock \showarticletitle{Frequency domain acceleration of convolutional
  neural networks on CPU-FPGA shared memory system}. In
  \bibinfo{booktitle}{\emph{Proceedings of the 2017 ACM/SIGDA International
  Symposium on Field-Programmable Gate Arrays}}. ACM, \bibinfo{pages}{35--44}.
\newblock


\bibitem[\protect\citeauthoryear{Zhang, Sun, Fang, Zhou, Pan, and Cong}{Zhang
  et~al\mbox{.}}{2018a}]%
        {2018_TCAD_Caffine}
\bibfield{author}{\bibinfo{person}{Chen Zhang}, \bibinfo{person}{Guangyu Sun},
  \bibinfo{person}{Zhenman Fang}, \bibinfo{person}{Peipei Zhou},
  \bibinfo{person}{Peichen Pan}, {and} \bibinfo{person}{Jason Cong}.}
  \bibinfo{year}{2018}\natexlab{a}.
\newblock \showarticletitle{Caffeine: Towards uniformed representation and
  acceleration for deep convolutional neural networks}.
\newblock \bibinfo{journal}{\emph{IEEE Transactions on Computer-Aided Design of
  Integrated Circuits and Systems}} (\bibinfo{year}{2018}).
\newblock


\bibitem[\protect\citeauthoryear{Zhang, Ye, Zhang, Tang, Wen, Fardad, and
  Wang}{Zhang et~al\mbox{.}}{2018b}]%
        {2018_ECCV_ADMMpruning}
\bibfield{author}{\bibinfo{person}{Tianyun Zhang}, \bibinfo{person}{Shaokai
  Ye}, \bibinfo{person}{Kaiqi Zhang}, \bibinfo{person}{Jian Tang},
  \bibinfo{person}{Wujie Wen}, \bibinfo{person}{Makan Fardad}, {and}
  \bibinfo{person}{Yanzhi Wang}.} \bibinfo{year}{2018}\natexlab{b}.
\newblock \showarticletitle{A systematic dnn weight pruning framework using
  alternating direction method of multipliers}. In
  \bibinfo{booktitle}{\emph{Proceedings of the European Conference on Computer
  Vision (ECCV)}}. \bibinfo{pages}{184--199}.
\newblock


\bibitem[\protect\citeauthoryear{Zhou, Niu, and Zhang}{Zhou
  et~al\mbox{.}}{2019}]%
        {Lowrank_IEEE_2019}
\bibfield{author}{\bibinfo{person}{Wei Zhou}, \bibinfo{person}{Yue Niu}, {and}
  \bibinfo{person}{Guanwen Zhang}.} \bibinfo{year}{2019}\natexlab{}.
\newblock \showarticletitle{Sensitivity-oriented layer-wise acceleration and
  compression for convolutional neural network}.
\newblock \bibinfo{journal}{\emph{IEEE Access}}  \bibinfo{volume}{7}
  (\bibinfo{year}{2019}), \bibinfo{pages}{38264--38272}.
\newblock


\end{thebibliography}










\end{document}